\documentclass[aps,pra,showpacs,twocolumn,superscriptaddress,floatfix]{revtex4-2}
\usepackage{graphicx}
\usepackage{dcolumn}
\usepackage{bm}
\usepackage{hyperref}
\usepackage{comment}
\usepackage{xcolor}   
\usepackage{amsmath}
\usepackage{float}
\usepackage{amssymb}   
\usepackage{physics}

\usepackage{amsfonts}

\usepackage{placeins}
\usepackage{array}
\usepackage{amsmath}
\usepackage{tabularx}

\begin{document}

\preprint{APS/123-QED}

\title{Which-crystal information and wave–particle duality in induced-coherence interferometry
}

\newcommand{\IISc}{Quantum Optics and Quantum Information, 
Department of Electronic and Systems Engineering, 
Indian Institute of Science, Bengaluru 560012, India.}

\author{L. Theerthagiri  }
\email{tgiri9157@gmail.com}
\affiliation{\IISc}

\date{\today}

\begin{abstract}
We provide an operational reinterpretation of wave–particle complementarity in the low-gain Zou–Wang–Mandel (ZWM) induced-coherence interferometer. In the low gain limit, each photon pair is emitted by either one of two nonlinear crystals. Preparing nonorthogonal conditional idler states that encode which-crystal information. While previous studies inferred distinguishability indirectly from signal visibility with undetected idler photons. We show that the idler states naturally define a binary quantum hypothesis-testing problem. By performing optimal measurements on the idler, we analyze this task using both zero-error measurement unambiguous discrimination (Ivanovic–Dieks–Peres (IDP)) and minimum-error discrimination (Helstrom bound). We show that the signal visibility equals the optimal inconclusive probability of unambiguous discrimination. The Helstrom bound gives the optimal probability of identifying the emitting crystal.
While signal visibility is an ensemble-averaged expectation value, the IDP and Helstrom strategies correspond to optimal single-photon decision measurements on the idler. The decision problem concerns inferring a past source event from a present measurement outcome.
This establishes wave–particle duality in induced coherence as a manifestation of optimal quantum decision strategies rather than a purely geometric constraint. We further extend the analysis to the presence of thermal photons introduced in the object arm, which render the conditional idler states mixed. In this case, both the visibility and the achievable distinguishability are reduced, reflecting the fundamental limitations imposed by mixed-state discrimination. The approach is model-independent and applies to general two-path interferometers with markers.
\end{abstract}

\maketitle

\section{Introduction}
Interferometry based on path indistinguishability was introduced in the induced-coherence experiments of Zou, Wang, and Mandel (ZWM)~\cite{wang1991induced,zou1991induced,Zou1992}.
In these experiments, indistinguishable idler modes erase which-way information and induce first-order coherence between spatially separated signal fields. Single-photon interference then appears in the signal arm, even when the idler photon is not detected~\cite{Shafiee2024}. This regime provides a natural platform for probing the quantum origin of induced coherence.

The quantum versus classical character of induced coherence has been the subject of active debate. Contextuality-based analyses identify measurement settings in
which induced coherence leads to predictions incompatible with noncontextual hidden-variable models~\cite{Shafiee2024}. However, classical models reproduce several interference features of SPDC, including induced-coherence visibility in the
low-gain regime~\cite{Kulkarni2022}. These results indicate that interference visibility alone does not fully characterize which-way or which-source information in induced-coherence interferometry.

The visibility \(V_{Signal}\) of the signal interference fringes and the which-way distinguishability \(D_{Signal}\) are linked by the complementarity relation \(D^{2}+V^{2}=1\) for pure two-path markers~\cite{Wootters1979,Englert1996,Jaeger1993}. By contrast, Englert’s relation $D^{2}+V^{2}\le 1$ formulates wave--particle complementarity in terms of the overlap of which-path marker states~\cite{Englert1996}. Bagan \textit{et al.}~\cite{Bagan} further developed operational duality relations using discrimination games defined on abstract quantum states. However, the operational meaning of wave–particle complementarity in induced-coherence interferometry, particularly at the level of optimal single-photon inference, remains unclear.

A natural framework for addressing this limitation is provided by quantum decision theory. In particular, existing induced-coherence experiments do not access the full
which-crystal information encoded in the idler state.
 In the low-gain ZWM interferometer, each photon pair originates from either crystal~A or crystal~B, but never both. This defines a binary quantum hypothesis test \cite{Helstrom}: which crystal emitted the photon pair. The corresponding conditional idler states are generally nonorthogonal, so the previous discrimination is fundamentally constrained by quantum mechanics.

Optimal discrimination of nonorthogonal quantum states has been widely studied. Earlier experiments demonstrated the Ivanovic–Dieks–Peres (IDP) unambiguous discrimination measurement using optical interferometers~\cite{Clarke2001IDP,Herzog2005MixedUSD}. Related works have investigated path information and the past of photons inside interferometers using weak or frequency-based marking techniques~\cite{Zhou2017,Danan2013Trajectories}, as well as unambiguous path discrimination in Mach–Zehnder interferometers~\cite{Len2018IDP}. These studies address questions of which-path information or spatial presence.
In contrast, the present work focuses on induced-coherence interferometry and addresses a different retrodictive task: identifying which physical source emitted the photon pair.

In this work, we adopt a decision-theoretic perspective and analyze induced coherence using quantum state discrimination on the idler. For pure markers, we give the equality $D^{2}+V^{2}=1$ a direct \emph{operational} meaning. We identify the fringe visibility $V_{\mathrm{Signal}}$ with the optimal
inconclusive probability of the Ivanovic--Dieks--Peres (IDP) unambiguous discrimination measurement (USD), $P_{inc}^{\mathrm{opt}}$.
We show that the ZWM geometry itself implements the optimal three-outcome IDP POVM~\cite{Ivanovic,Dieks,Peres,Barnett}.
No ancillary modes or adaptive measurements are required. We also consider the complementary minimum-error strategy given by the Helstrom bound. This measurement maximizes the probability of correctly identifying which crystal emitted the photon pair.
The resulting distinguishability quantifies which-crystal information in the minimum-error sense. Together, the IDP and Helstrom strategies yield complementary operational
interpretations of wave--particle duality in induced coherence.

\begin{figure}[t]
    \centering
    \includegraphics[width=0.4\textwidth]{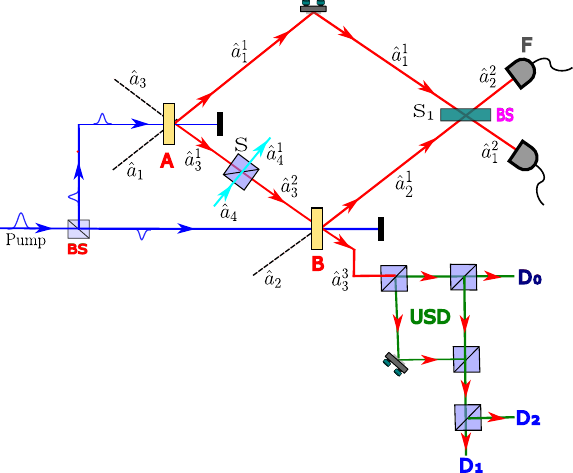}
    \caption{
    Implementation of optimal unambiguous state discrimination (USD)~\cite{Barnett,Clarke2001IDP,Len2018IDP} in the induced-coherence (Zou--Wang--Mandel) interferometer~\cite{Torres2024}.
    Two coherently pumped nonlinear crystals (A and B) generate
    signal--idler photon pairs in the low-gain regime. Each emission event originates from either crystal~A or crystal~B but never both. The signal modes interfere at a balanced beam splitter $S_1$, producing single-photon fringes with visibility $V$.
    The idler modes ($\hat{a}_{3}^{2},\hat{a}_{3}^{3}$) form the two nonorthogonal marker states
    $|\phi_A\rangle$ and $|\phi_B\rangle$, whose overlap determines $V$. The output of the first idler ($\hat{a}_{3}^{2}$) beam splitter is directed to detector $D_0$, realizing the inconclusive IDP outcome $\Pi_{inc}$. The detectors $D_1$ and $D_2$ project onto the orthogonal complement states $|\phi_B^{\perp}\rangle$ and
    $|\phi_A^{\perp}\rangle$, yielding the conclusive IDP outcomes $\Pi_A$ and $\Pi_B$. 
   }
    \label{fig:ZWM_IDP}
\end{figure}

\section{Induced-coherence interferometer marker states and complementarity }\label{marker}

In the ZWM setup as shown in Fig.~\ref{fig:ZWM_IDP}, coherently pump the two nonlinear crystals. In the low-gain regime~\cite{Shafiee2024,boyd,Kolobov2017,giri1}, each emission event corresponds to one of two mutually exclusive hypotheses. These hypotheses are encoded in the conditional idler states prepared by crystals A and B, which are generally nonorthogonal. As a result, perfect discrimination is impossible.  Any zero-error identification must rely on an unambiguous state-discrimination (USD) strategy.

In the low-gain regime, the joint state generated by two coherently pumped SPDC crystals can be written schematically as
\begin{equation}
|\Psi\rangle_{total} \approx |0\rangle
+ g_A\,|1_{S_A},1_{I_A}\rangle
+ g_B\,|1_{S_B},1_{I_B}\rangle,
\label{twocrytal}
\end{equation}
with small pair-creation amplitudes \(g_A\) and \(g_B\). When the idler modes  (\(\hat{a}_{3}^{1}=I_A\)) are aligned to be path-identical, \(\hat{a}_{3}^{2}=I_{A}^{'}\) and \(\hat{a}_{3}^{3}=I_B\)  become a single physical mode \(I_{A}^{'}=I_B=I\), and the biphoton component factorizes as
\begin{equation}
\begin{split}
g_A|1_{S_A},1_{I_A}\rangle + g_B|1_{S_B},1_{I_B}\rangle
\;\longrightarrow\;
\\[2pt]
|1_I\rangle \otimes
\big( g_A |1_{S_A},0_{S_B}\rangle
+ g_B |0_{S_A},1_{S_B}\rangle \big),
\end{split}
\label{signalmode}
\end{equation}
so that the signal field is left in a coherent superposition even when the idler is not detected.

We now introduce an object in the shared idler $I$ path between the two crystals, modeled as a lossless beam splitter with amplitude transmittance \(t\) and reflectivity \(r\),
\(|r|^{2}+|t|^{2}=1\). In the single-pair sector, the biphoton state at the output of the two crystals can be written (see Appendix~\ref{app:one}) as
\begin{equation}
\begin{split}
|\psi\rangle_{SI} = \frac{1}{\sqrt{2}}\Big[
&|1_{S_A},0_{S_B}\rangle
\big(r|1_{I_{A}^{'}},0_{I_B}\rangle+t|0_{I_{A}^{'}},1_{I_B}\rangle\big)
\\
&+ e^{i\phi}|0_{S_A},1_{S_B}\rangle
|0_{I_{A}^{'}},1_{I_B}\rangle
\Big],
\end{split}
\label{eq:ZWM_state}
\end{equation}
where \(\phi\) is the relative pump phase. Conditioned on emission in
crystal~A (the \(|1_{S_A},0_{S_B}\rangle\) branch), the idler state is
\begin{equation}
|\phi_A\rangle = r\,|1_{I_{A}^{'}},0_{I_B}\rangle + t\,|0_{I_{A}^{'}},1_{I_B}\rangle,
\end{equation}
while emission in crystal~B (the \(|0_{S_A},1_{S_B}\rangle\) branch)
leaves the idler in
\begin{equation}
|\phi_B\rangle = |0_{I_{A}^{'}},1_{I_B}\rangle.
\end{equation}

Tracing over the idler and recombining \(S_1\) on a balanced
signal beam splitter gives single-photon interference in the signal arm $F$ with visibility
\begin{equation}
V = |\langle\phi_A|\phi_B\rangle| = |t|.
\end{equation}
A natural measure of which-path distinguishability for the pure idler
alternatives is
\begin{equation}
D = \sqrt{1-|\langle\phi_A|\phi_B\rangle|^{2}} = |r|.
\end{equation}
Thus the ZWM interferometer realizes the standard complementarity
relation
\begin{equation}
D^{2} + V^{2} = 1.
\label{dv}
\end{equation}

An extension to imperfect idler mode overlap is given in
Appendix~\ref{app:two} and leads to the reduction
\(V \rightarrow |\gamma|\,|t|\),
with an unchanged complementarity relation $D^{2}+V^{2}=1$ in the pure-state model.

 In Eq.~\eqref{signalmode}, signal written as a coherent superposition of two emission alternatives. The path identical idlers $I$ are creating a two assumption, with object and without object. 

If object $S$ is not present, idler modes (\(\hat{a}_{3}^{3}=I_A\))  associated with the two crystals are rendered path-identical $I$, the conditional idler states are identical, $|\phi_A\rangle = |\phi_B\rangle$.
Tracing over the idler degree of freedom then yields a reduced signal state with a non-vanishing off-diagonal coherence term. Resulting in maximal first-order interference in the signal arm.
This is the basic mechanism of induced coherence in the ZWM interferometer. On the other hand, the idler modes remain distinguishable and occupy orthogonal spatial modes. The conditional idler states are orthogonal, $\langle \phi_A | \phi_B \rangle = 0$. In this case, tracing over the idler removes all coherence between the two signal emission alternatives, and no first-order interference is observed. Thus, path identity of the idler modes is a necessary condition for induced coherence.

If the object $S$ present, the common idler $I$ path modifies the idler associated with only one emission alternative.
Although the idler modes remain path-identical. The object renders the conditional idler states nonorthogonal,
$0 < |\langle \phi_A | \phi_B \rangle| < 1$.
After tracing over the idler, the reduced density matrix of the signal contains an off-diagonal coherence term proportional to this overlap,
\begin{equation}
\rho_{S_A ,S_B} \propto \langle \phi_A | \phi_B \rangle .
\end{equation}

As a result, the signal exhibits partial induced coherence with a visibility determined by the degree of nonorthogonality of the idler states.

If the idler modes $I$ are not path-identical. The conditional idler states remain orthogonal regardless of the presence of the object, and induced coherence is completely suppressed.
Thus, path identity enables induced coherence, while the object controls its degradation by introducing controlled nonorthogonality between the conditional idler states.


In the induced-coherence interferometer, the signal photons
$(\hat{a}_1^{\,1}=S_A,\ \hat{a}_2^{\,1}=S_B)$ propagate in two orthogonal spatial modes [see Eq.~\eqref{signalmode}]. So, which-path information is well defined for the signal degree of freedom. By contrast, the idler modes are deliberately aligned to be path-identical, rendering which-path
information ill defined for the idler. The only physically meaningful alternative for the idler is therefore the emission origin of the photon pair. By making the idler modes produced by different crystals identical, the emission origin of the photon pair becomes indistinguishable, which induces first-order coherence between the corresponding signal modes.If the pair is generated in crystal~A, the idler photon interacts with the object, whereas if it is generated in crystal~B, the idler bypasses the object. Consequently, the relevant wave--particle tradeoff for the idler concerns a retrodictive inference between two mutually exclusive emission events, encoded in a pair of nonorthogonal conditional idler states.

Previous ZWM experiments have measured visibility and which-path
information as evidence of the complementarity relation
\(D^{2}+V^{2}=1\) \cite{Torres2024}, but they do not address the fundamental limit on identifying the emitting crystal in a single-shot measurement. This motivates the operational question: \emph{How well can one unambiguously (error-free) discriminate which
crystal produced the photon pair?} Because the conditional idler states are nonorthogonal, perfect identification is impossible, and any zero-error strategy must rely on unambiguous discrimination.

\section{Quantum decision strategies for which-crystal discrimination}\label{hypo}

In the ZWM setup, low-gain regime induced-coherence scenario, the idler subsystem encodes the binary hypothesis \emph{which crystal fired}, more details see Appendix~\ref{app:two}. We denote the two hypotheses by

\begin{equation}
    \begin{split}
        \mathcal{H}_A &: \ \hat{\rho} = \hat{\rho}_A \ \ \text{(crystal A fired)}, \qquad \text{prior } p_0,\\ 
\mathcal{H}_B &: \ \hat{\rho} = \hat{\rho}_B \ \ \text{(crystal B fired)}, \qquad \text{prior } p_1,
    \end{split}
\end{equation}
with $p_0,p_1\ge 0$ and $p_0+p_1=1$.

In the idler mode $I$ performs a dichotomous POVM $\{\Pi_A,\Pi_B\}$ with outcomes $\mu\in\{0,1\}$ satisfying
\begin{equation}
\{\Pi_A,\Pi_B\}, \qquad \Pi_A \ge 0,\ \Pi_B \ge 0,\ \Pi_A + \Pi_B = \openone,
\label{povm}
\end{equation}
where outcome $\Pi_A$ corresponds to the decision ``A fired'' ($\mu =0$) and outcome $\Pi_B$ corresponds to the
decision ``B fired'' ($\mu =1$).

In the Conditional error probabilities,
If $\mathcal{H}_A$ is true (state $\hat{\rho}_A$ was prepared), an error occurs when crystal outputs ``B'',
i.e., when outcome $\Pi_B$ clicks. Similarly, if $\mathcal{H}_B$ is true (state $\hat{\rho}_B$ was prepared), an error occurs when crystal outputs ``A'', i.e., when outcome $\Pi_A$ clicks:

\begin{equation}
\begin{split}
   P_A &:= P(H_1|H_0)
     = \Pr(\mu=1\,|\,\hat{\rho}=\hat{\rho}_A)
     = \Tr(\Pi_B \hat{\rho}_A),\\
P_B &:= P(H_0|H_1)
     = \Pr(\mu=0\,|\,\hat{\rho}=\hat{\rho}_B)
     = \Tr(\Pi_A \hat{\rho}_B). 
     \label{eq:PM_def} 
\end{split}
\end{equation}

The Bayes average error probability of which crystal fired is
\begin{equation}
P_{\rm err} = p_0 P_A + p_1 P_B
            = p_0\,\Tr(\Pi_B\hat{\rho}_A) + p_1\,\Tr(\Pi_A\hat{\rho}_B).
\label{eq:Perr_bayes}
\end{equation}

\section{Minimum-error strategy: Helstrom bound}\label{mini}

In the low-gain ZWM interferometer, the two hypotheses ``crystal A fired'' and ``crystal B fired'' are encoded in the conditional idler states see more details Appendix~\ref{app:three}
\begin{equation}
\ket{\phi_A}= r\ket{1,0}+t\ket{0,1},\qquad
\ket{\phi_B}= \ket{0,1},
\label{eq:ZWM_idler}
\end{equation}
defined on the single-photon idler subspace spanned by
$\{\ket{1,0},\ket{0,1}\}$, with $|r|^2+|t|^2=1$ and overlap
\begin{equation}
s:=|\braket{\phi_A}{\phi_B}|=|t|.
\label{eq:s}
\end{equation}
We now derive the optimal minimum-error (Helstrom) discrimination
strategy between $\rho_A=\ket{\phi_A}\bra{\phi_A}$ and
$\rho_B=\ket{\phi_B}\bra{\phi_B}$, and we identify the optimal POVM explicitly.

 Helstrom operator and minimum error probability, For priors $p_1,p_2\ge 0$ with $p_1+p_2=1$, define the Helstrom
operator
\begin{equation}
\Delta := p_1\rho_A-p_2\rho_B.
\label{eq:Helstrom_operator_def}
\end{equation}
The Helstrom theorem states that the minimum achievable average error probability is
\begin{equation}
P_{err}^{\min}=\frac12\Big(1-\|\Delta\|_1\Big),
\label{eq:Helstrom_error_general}
\end{equation}
where $\|\cdot\|_1$ denotes the trace norm, and that the optimal
two-outcome POVM is the projective measurement onto the positive and negative eigenspaces of $\Delta$, see more details Appendix~\ref{app:three}.

The eigenvalues of $\Delta$ are therefore
\begin{equation}
\lambda_\pm=\pm\frac12\sqrt{a^2+|b|^2}
=\pm\frac{|r|}{2}.
\label{eq:Delta_eigs}
\end{equation}
Since $\Delta$ is Hermitian, its trace norm is the sum of absolute
eigenvalues,
\begin{equation}
\|\Delta\|_1 = |\lambda_+|+|\lambda_-|=|r|.
\label{eq:trace_norm_Delta}
\end{equation}
Substituting Eq.~\eqref{eq:trace_norm_Delta} into Eq.~\eqref{eq:Helstrom_error_general}
yields the Helstrom minimum error and maximum success probabilities:
\begin{equation}
P_{err}^{\min}=\frac12(1-|r|),\qquad
P_{suc}^{\max}=1-P_{err}^{\min}=\frac12(1+|r|).
\label{eq:Helstrom_error_success_equalpriors}
\end{equation}
In terms of the overlap in Eq.~\eqref{eq:s} $s=|t|$ we have $|r|=\sqrt{1-|t|^2}=\sqrt{1-s^2}$, so
\begin{equation}
P_{err}^{\min}
=\frac12\Big(1-\sqrt{1-s^2}\Big)
=\frac12\Big(1-\sqrt{1-|t|^2}\Big).
\label{eq:Helstrom_error_overlap_form}
\end{equation}

It is natural to define the (equal-prior) Helstrom trace-distance
distinguishability as (See more details in Appendix.~\ref{app:four})
\begin{equation}
D_{\mathrm{Hel}}:=\|\Delta\|_1=|r|=\sqrt{1-|t|^2}.
\label{eq:DHel_def}
\end{equation}

Since the ZWM singles visibility is $V=|t|$, one obtains the saturated pure-marker duality relation
\begin{equation}
D_{\mathrm{Hel}}^2+V^2=|r|^2+|t|^2=1.
\label{eq:Helstrom_complementarity}
\end{equation}
Thus, in the pure-marker regime, the distinguishability appearing in the complementarity relation coincides with the Helstrom distinguishability for minimum-error identification of the emitting crystal.

In the optimal Helstrom POVM (explicit eigenmodes). The optimal Helstrom POVM in Eq.~\eqref{povm} for equal priors is the projective measurement onto the eigenvectors of $\Delta$:
\begin{equation}
\Pi_A=\ket{\omega_+}\bra{\omega_+},\quad
\Pi_B=\ket{\omega_-}\bra{\omega_-},
\quad
\Pi_A+\Pi_B=\openone.
\label{eq:Helstrom_POVM_def}
\end{equation}

A convenient normalized eigenbasis of $\Delta$ is then
\begin{equation}
\begin{split}
   \ket{\omega_+}
&=
\cos\frac{\theta}{2}\,\ket{1,0}
+
e^{-i\varphi}\sin\frac{\theta}{2}\,\ket{0,1},
\label{eq:omega_plus}
\\[3pt]
\ket{\omega_-}
&=
\sin\frac{\theta}{2}\,\ket{1,0}
-
e^{-i\varphi}\cos\frac{\theta}{2}\,\ket{0,1},
\end{split}
\end{equation}
with $\braket{\omega_+}{\omega_-}=0$.
The decision rule is: a detection in the $\ket{\omega_+}$ output mode corresponds to guessing $H_1$ (crystal A), while detection in the $\ket{\omega_-}$ output mode corresponds to guessing $H_2$ (crystal B). Physically, this Helstrom measurement is implemented by a single two-mode interferometer acting on the idler outputs in Eq.~\eqref{eq:ZWM_idler} ($\vert \phi_A\rangle,\vert \phi_B\rangle$) with controllable relative phase $\varphi$ and mixing angle $\theta/2$, followed by two photon detectors.

For completeness, for arbitrary priors $p_1\neq p_2$, the Helstrom bound for two pure states with overlap in Eq.~\eqref{eq:s} $s=|\braket{\phi_A}{\phi_B}|$ is
\begin{equation}
P_{err}^{\min}=\frac12\Big(1-\sqrt{1-4p_1p_2\,s^2}\Big)
=\frac12\Big(1-\sqrt{1-4p_1p_2\,|t|^2}\Big).
\label{eq:Helstrom_error_unequalpriors}
\end{equation}
Equivalently,
\begin{equation}
\|\Delta\|_1=\sqrt{1-4p_1p_2\,|t|^2}.
\label{eq:Helstrom_norm_unequalpriors}
\end{equation}
For the equal-pump ZWM scenario $p_1=p_2=\tfrac12$, these expressions reduce to Eqs.~\eqref{eq:Helstrom_error_success_equalpriors}--\eqref{eq:DHel_def}.

Equation~\eqref{eq:Helstrom_complementarity} shows that, for pure idler marker states, wave--particle duality in the ZWM interferometer is nothing but the geometry of minimum-error quantum state discrimination. The visibility \(V\) quantifies
the wave character through the overlap of the two marker states. However, \(D_{\mathrm{Hel}}\) quantifies the particle character as the optimal Helstrom distinguishability---i.e., the best possible ability to identify which crystal fired with minimum average error.

Importantly, this interpretation is operational: \(D_{\mathrm{Hel}}\) is not an abstract distinguishability parameter but directly determines the optimal
decision strategy. The error probability via the Helstrom bound
\eqref{eq:Helstrom_error_V}. When the marker states are pure, the trade-off between interference visibility and which-crystal information is saturated. In the presence of loss or mixing (mixed marker states), the equality generally becomes an inequality, \(D_{\mathrm{Hel}}^2 + V^2 \le 1\), reflecting reduced operational distinguishability (See Appendix~\ref{nine}).

\begin{figure}
  \centering
  \includegraphics[width=0.7\linewidth]{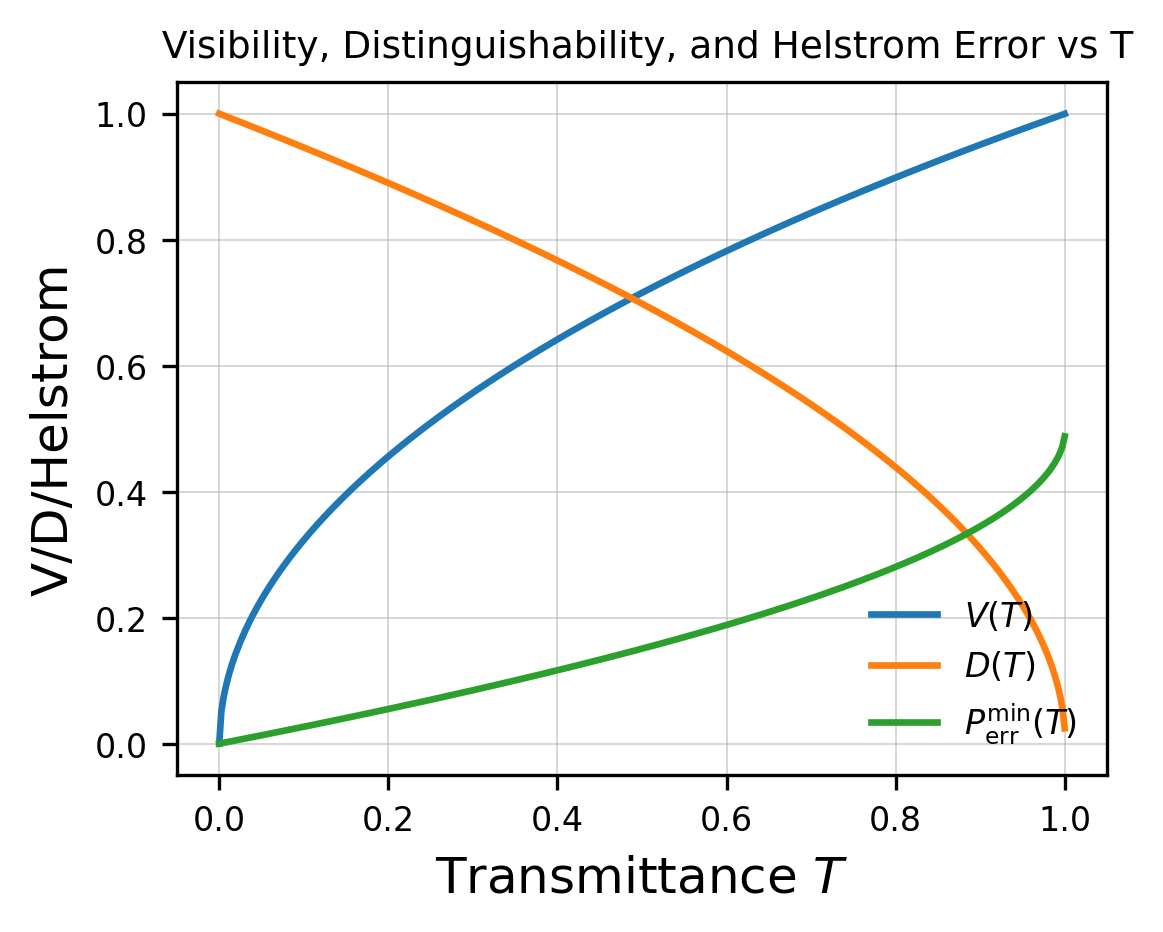}
 \caption{Visibility $V(T)$, distinguishability
$D(T)=\sqrt{1-V(T)^2}$, and the Helstrom minimum error probability $P_{\mathrm{err}}^{\min}(T)=\tfrac{1}{2}\big[1-D(T)\big]$ as functions of the object transmittance $T$ in the induced-coherence interferometer.
Increasing transmittance enhances signal interference visibility while reducing which-crystal distinguishability encoded in the idler. The Helstrom bound quantifies the optimal minimum-error probability for inferring which nonlinear crystal emitted the photon pair from a single idler measurement.
Illustrating the operational complementarity between interference and source discrimination.}
  \label{fig2}
\end{figure}

In Fig.~\ref{fig2}, the Helstrom error probability is transparent in the two extreme limits of the object transmittance. For an opaque object ($T=0$), the conditional idler states associated with emission from crystals~A and~B are orthogonal. If crystal~A fires the idler is reflected, whereas if crystal~B fires the idler bypasses the object.
A single idler detection event therefore identifies the emitting crystal with certainty. The minimum error probability vanishes, $P_{\mathrm{err}}^{\min}=0$. Conversely, for a fully transparent object ($T=1$), both emission alternatives prepare the same idler state. The idler then carries no information about the source. Any measurement strategy reduces to random guessing, yielding $P_{\mathrm{err}}^{\min}=1/2$.
Intermediate transmittance values interpolate smoothly between these limits ($0< T< 1$), quantifying how partial which-crystal information degrades induced coherence.

\section{Zero-error strategy: IDP state discrimination}
\label{IDP_ZWM}

In this section,  we treat the same ``which-crystal fired?'' problem, but now in the \emph{unambiguous} (zero-error) discrimination setting \cite{Barnett,Clarke2001IDP,Len2018IDP}: whenever the measurement makes a conclusive guess, it is \emph{never wrong}, at the price of allowing an \emph{inconclusive} outcome (See Appendix~\ref{five}).

In the Sec.\ref{marker} and Sec.~\ref{hypo}, two pure idler marker states are nonorthogonl, perfect identification is impossible. In the Idler measurement, any zero-error strategy must rely on unambiguous discrimination (USD) as shown in the Fig.~\ref{fig:ZWM_IDP}. The optimal solution is the IDP measurement, whose minimal inconclusive probability is set by the state overlap.

The idler states $|\phi_A\rangle$ and $|\phi_B\rangle$ produced by the two crystals occupy only the single-photon idler subspace spanned by $|1,0\rangle$ and $|0,1\rangle$.  In this two-dimensional subspace, each $|\phi_j\rangle$ admits a unique (up to phase) orthogonal complement $|\phi_j^\perp\rangle$.

We use the parametrization
\begin{equation}
|\phi_A\rangle = r\,|1,0\rangle + t\,|0,1\rangle, \qquad
|\phi_B\rangle = |0,1\rangle,
\end{equation}
with $|r|^2 + |t|^2 = 1$.

The orthogonal complement of $|\phi_B\rangle$ is immediate:
\begin{equation}
|\phi_B^\perp\rangle = |1,0\rangle.
\end{equation}

For $|\phi_A\rangle$, we write a general state in this subspace as
$|\phi_A^\perp\rangle = \alpha\,|1,0\rangle + \beta\,|0,1\rangle$ and impose
$\langle \phi_A^\perp | \phi_A\rangle = 0$, giving
\begin{equation}
\alpha^* r + \beta^* t = 0.
\end{equation}
A convenient normalized solution is
\begin{equation}
|\phi_A^\perp\rangle
= t^*\,|1,0\rangle - r^*\,|0,1\rangle,
\end{equation}
where we used $|r|^2+|t|^2=1$.

An unambiguous discrimination measurement is a 3-outcome POVM \cite{Barnett,Peres,Dieks,Clarke2001IDP}
\begin{equation}
\{\Pi_A,\Pi_B,\Pi_{inc}\},\quad \Pi_A,\Pi_B,\Pi_{inc}\ge 0,\quad
\Pi_A+\Pi_B+\Pi_{inc}=\openone,
\end{equation}
with the decision rule:
\begin{itemize}
\item outcome \(\Pi_A\): declare ``state \(\ket{\phi_A}\)'' (crystal A fired),
\item outcome \(\Pi_B\): declare ``state \(\ket{\phi_B}\)'' (crystal B fired),
\item outcome \(\Pi_{inc}\): declare ``Inconclusive ($inc$)''.
\end{itemize}

Zero-error (unambiguous) constraints, unambiguous discrimination requires that a conclusive outcome never clicks for the wrong state:
\begin{equation}
\Tr(\Pi_A\rho_B)=0,\qquad \Tr(\Pi_B\rho_A)=0,
\qquad \rho_j=\ket{\phi_j}\bra{\phi_j}.
\label{eq:IDP_zero_error}
\end{equation}

Since \(\rho_j\) are rank-1 projectors and \(\Pi_k\ge 0\), Eq.~\eqref{eq:IDP_zero_error} is equivalent to the \emph{support} conditions

\begin{equation}
\Pi_A\ket{\phi_B}=0,\qquad \Pi_B\ket{\phi_A}=0.
\end{equation}

In a two-dimensional Hilbert space, each condition forces \(\Pi_A\) (resp. \(\Pi_B\)) to be supported on the one-dimensional subspace orthogonal to \(\ket{\phi_B}\)
(resp. \(\ket{\phi_A}\)). Therefore we may write

\begin{equation}
\Pi_A = a\,\ket{\phi_B^\perp}\bra{\phi_B^\perp},\qquad
\Pi_B = b\,\ket{\phi_A^\perp}\bra{\phi_A^\perp},
\label{eq:IDP_rank1_form}
\end{equation}
with parameters \(a,b\ge 0\), and where

\begin{equation}
\braket{\phi_B^\perp}{\phi_B}=0,\qquad \braket{\phi_A^\perp}{\phi_A}=0,
\qquad \|\phi_A^\perp\|=\|\phi_B^\perp\|=1.
\end{equation}

The inconclusive POVM element is then fixed by completeness:
\begin{equation}
\Pi_{inc}=\openone-\Pi_A-\Pi_B.
\label{eq:Pi_question_def}
\end{equation}

The only remaining requirement is positivity:
\begin{equation}
\Pi_{inc}\ge 0 \quad \Longleftrightarrow \quad \Pi_A+\Pi_B \le \openone.
\label{eq:positivity_constraint}
\end{equation}

Success and inconclusive probabilities. For state \(\ket{\phi_A}\), the probability of a correct conclusive identification is
\begin{equation}
P(A|A)=\Tr(\Pi_A\rho_A)=a\,|\braket{\phi_B^\perp}{\phi_A}|^2,
\end{equation}
and for state \(\ket{\phi_B}\),
\begin{equation}
P(B|B)=\Tr(\Pi_B\rho_B)=b\,|\braket{\phi_A^\perp}{\phi_B}|^2.
\end{equation}
The inconclusive probabilities are
\begin{equation}
\begin{split}
Q_A\equiv & P(inc|A)=1-P(A|A),\\
Q_B \equiv & P(inc|B)=1-P(B|B),
\end{split}
\end{equation}
and the \emph{average} inconclusive probability is
\begin{equation}
Q \equiv P(inc) = p_1 Q_A + p_2 Q_B.
\label{eq:Q_def}
\end{equation}
The IDP problem is: \emph{minimize \(Q\) subject to \(\Pi_{inc}\ge 0\).}

In the Schr\"odinger-picture derivation see more details Appendix~\ref{six}. Unambiguous discrimination between $|\phi_A\rangle$ and $|\phi_B\rangle$ with equal priors requires a three-outcome POVM $\{\Pi_A,\Pi_B,\Pi_{inc}\}$. The corresponding POVM, to conclusive identification of $|\phi_A\rangle$, conclusive identification of $|\phi_B\rangle$, and an inconclusive result. Physically, such a POVM can be realized by embedding the two-dimensional idler subspace spanned by $\{|1,0\rangle_{I_{A}^{'},I_B},|0,1\rangle_{I_{A}^{'},I_B}\}$ into a three-dimensional single-photon space. By introducing a third auxiliary mode $C$ initially in the vacuum state $|0_C\rangle$.

In the single-photon sector of the three modes $(I_{A}^{'},I_B,C)$ we use the basis
\begin{equation}
\begin{aligned}
  \{\,|100\rangle,\,|010\rangle,\,|001\rangle\,\}_{I_2,I_3,C}
  :=\;
  \{&
     |1,0,0\rangle,\\
     &|0,1,0\rangle,\\
     &|0,0,1\rangle
  \}_{I_{A}^{'},I_B,C}.
\end{aligned}
\end{equation}

The Ivanovic--Dieks--Peres (IDP) measurement is implemented by a three-mode unitary
$U_{\mathrm{IDP}}$ acting on the extended idler state
$|\phi_j\rangle\otimes|0_C\rangle$. The most general form of $U_{\mathrm{IDP}}$ consistent with error-free discrimination is
\begin{align}
  U_{\mathrm{IDP}}\,|\phi_1,0_C\rangle
  &= \sqrt{p_1}\,|100\rangle + \sqrt{q_1}\,|001\rangle,
  \label{eq:UIDP_phi1}\\
  U_{\mathrm{IDP}}\,|\phi_2,0_C\rangle
  &= \sqrt{p_2}\,|010\rangle + \sqrt{q_2}\,|001\rangle,
  \label{eq:UIDP_phi2}
\end{align}
where $p_j$ and $q_j$ ($j=1,2$) are nonnegative real numbers. In this representation

\begin{itemize}
  \item $p_1$ is the probability that input $|\phi_A\rangle$ produces a click in mode~1 (conclusive ``crystal~A''),
  \item $p_2$ is the probability that input $|\phi_B\rangle$ produces a click in mode~2 (conclusive ``crystal~B''),
  \item $q_1$ and $q_2$ are the probabilities that the respective inputs yield a click in mode~3 (inconclusive outcome).
\end{itemize}

The form of Eqs.~\eqref{eq:UIDP_phi1}–\eqref{eq:UIDP_phi2} enforces the zero-error conditions: for $|\phi_A\rangle$ there is no amplitude in $|010\rangle$ (the ``B'' detector never fires), and for $|\phi_B\rangle$ there is no amplitude in $|100\rangle$ (the ``A'' detector never fires).

The average success probability of the IDP measurement is then
\begin{equation}
  P_{suc}^{\mathrm{opt}}
  = \tfrac{1}{2}p_1 + \tfrac{1}{2}p_2
  = 1 - s,
\end{equation}
and the optimal inconclusive probability is
\begin{equation}
  P_{inc}^{\mathrm{opt}} = 1 - P_{suc}^{\mathrm{opt}} = s
  = |\langle\phi_A|\phi_B\rangle|.
\end{equation}

In the ZWM interferometer, $s = |\langle\phi_A|\phi_B\rangle| = |t| = V$, so the IDP theory yields
\begin{equation}
  P_{inc}^{\mathrm{opt}} = V,\qquad
  P_{suc}^{\mathrm{opt}} = 1 - V.
\end{equation}
Thus the conventional ZWM visibility is exactly the optimal failure probability of an error-free IDP measurement on the idler, while the which-path distinguishability is identified with the conclusive success probability. This provides the operational identity
\begin{equation}
  D^2 + \bigl(P_{inc}^{\mathrm{opt}}\bigr)^2 = 1
\end{equation}
for pure marker states, and establishes the interpretation $V = P_{inc}^{\mathrm{opt}}$ used in the main text.

\begin{figure}
  \centering
  \includegraphics[width=1\linewidth]{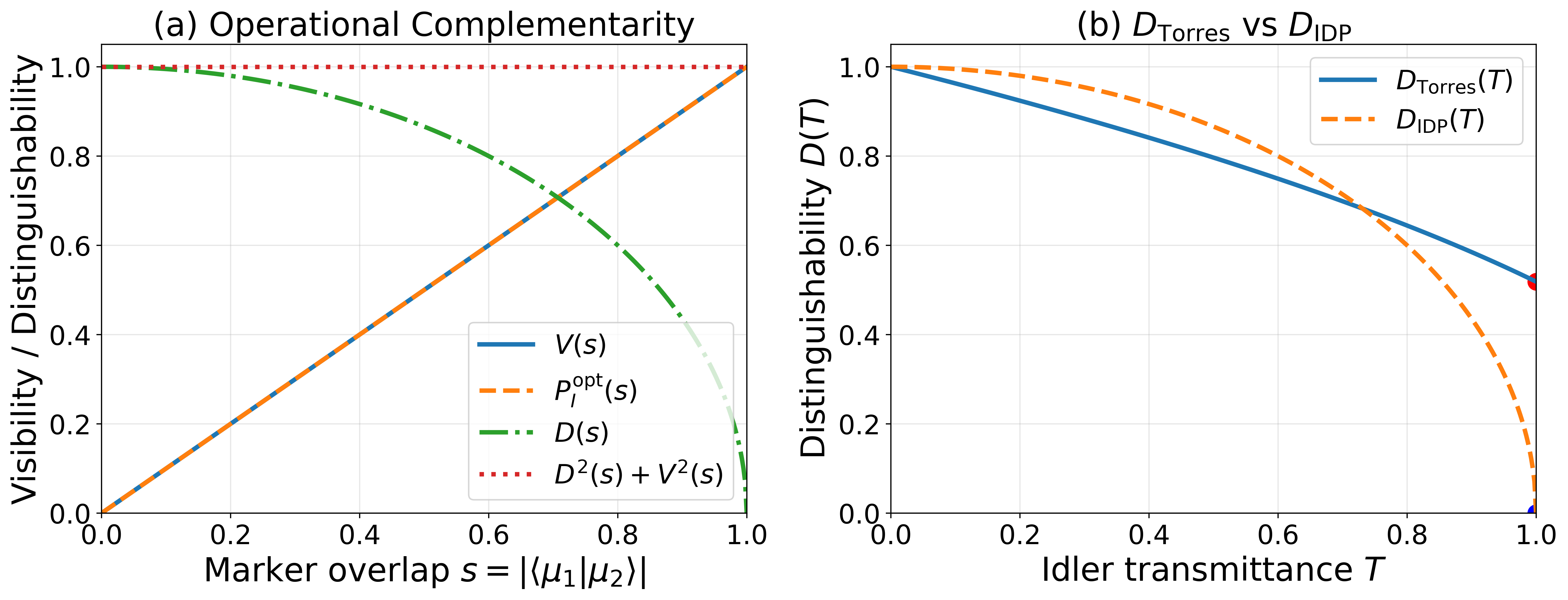}
 \caption{
(a) Operational complementarity between visibility $V$, distinguishability $D$, and the optimal IDP failure probability $P_{inc}^{\mathrm{opt}}$ for pure marker
states with overlap $s = |\langle\phi_A|\phi_B\rangle|$ (see Appendix~\ref{six}).
The curves satisfy $D^2+V^2=1$ and $V=P_{inc}^{\mathrm{opt}}$, identifying wave--particle duality with the optimality condition of unambiguous state discrimination.
(b) Comparison between distinguishability obtained via the coincidence-based method of Torres \textit{et al.}~\cite{Torres2024} (solid) and via optimal IDP discrimination on the idler (dashed). The coincidence method saturates at
$D_{\mathrm{coinc}}(T=1)\approx0.52$, whereas IDP extracts greater
distinguishability for intermediate $0<T<1$. Both approaches correctly give
$D=0$ at $T=1$, where the marker states become identical. The advantage of IDP
therefore appears only for partially overlapping markers.}

  \label{fig:TwoPanel_Comparison}
\end{figure}

Figure~\ref{fig:TwoPanel_Comparison} provides a direct operational comparison between distinguishability extracted from signal-arm visibility (Torres \textit{et al.}) \cite{Torres2024} and distinguishability obtained via optimal IDP discrimination performed on the idler. In panel (a), we plot the complementarity relation for pure which-crystal markers, showing that the visibility $\mathcal{V}$, the optimal IDP failure probability $P_{inc}^{\mathrm{opt}}$. The particle-like distinguishability $D$ are constrained by the operational identity $\mathcal{V}=P_{inc}^{\mathrm{opt}}$ and $D^2+\mathcal{V}^2=1$. This demonstrates that wave–particle duality is exactly equivalent to an optimal unambiguous state discrimination task for the which-path markers.

Panel (b) compares distinguishability extracted from the
coincidence-based method of Torres \textit{et al.} \cite{Torres2024} (solid curve). The distinguishability achievable via optimal Ivanovic--Dieks--Peres (IDP) unambiguous discrimination on the idler (dashed curve), both expressed in the same metric. In the Torres approach, $D$ is inferred from the imbalance between second-order correlations $g^{(2)}_{13}$ and $g^{(2)}_{23}$. It saturates at a finite value $D_{\mathrm{coinc}}(T=1) < 1$ even when the idler transmissivity $T$ is unity. In the opposite limit, the IDP strategy operates directly on the which-crystal idler marker states and achieves the optimal extraction of available which-path information. Both
methods converge to $D=0$ when $T=1$, since the idler marker states become identical and no measurement—IDP or otherwise—can distinguish the emission events. The advantage of IDP emerges in the intermediate regime $0 < T < 1$, where coincidence-based extraction underestimates the distinguishability available in principle. This highlights that path information in induced coherence is fundamentally limited by measurement strategy rather than interferometer geometry.

\section{General Marker state}
The same reasoning applies to any balanced two-path interferometer in the single-excitation regime with pure which-path markers $|\mu_A\rangle$ and $|\mu_B\rangle$. The fringe visibility is determined by their overlap (See Appendix~\ref{seven}),
\[
V = |\langle \mu_A | \mu_B \rangle|,
\]
and the corresponding distinguishability is
\[
D = \sqrt{1 - |\langle \mu_A | \mu_B \rangle|^{2}}.
\]
For equal priors, the optimal IDP strategy yields
$P_{inc}^{\mathrm{opt}} = |\langle \mu_A | \mu_B \rangle|$, so that
\[
D^2 + \big(P_{inc}^{\mathrm{opt}}\big)^2 = 1
\]
for any marker realization (polarization, spatial mode, time bin,
frequency, idler photon, etc.). See supplemental material for extensions to unequal priors and mode mismatch. In this sense, complementarity for pure markers is equivalent to the fundamental limit of a zero-error state-discrimination task.

\begin{table*}[t!]
\caption{Operational meaning of visibility and distinguishability as quantum decision tasks.}
\label{table}
\begin{ruledtabular}
\begin{tabular}{lcccc}
Strategy
& Outcomes
& Errors
& Operational quantity
& What $V$ means
\\ \hline
Previous ZWM setup
& $1$
& No errors
& $D=\sqrt{1-V^2}$
& $V$ (signal measurement)
\\
Helstrom (min-error)
& $2$
& Allowed (minimized)
& $D_{\mathrm{Hel}}=\tfrac12\|\rho_A-\rho_B\|_1$
& $V=|\langle\phi_A|\phi_B\rangle|$ (pure markers)
\\
USD / IDP (zero-error)
& $3$
& None (if conclusive)
& $D_{\mathrm{IDP}}=\sqrt{1-\bigl(P_{\mathrm{inc}}^{\mathrm{opt}}\bigr)^2}$
& $V=P_{\mathrm{inc}}^{\mathrm{opt}}$ (pure markers, equal priors)
\\
Mixed markers
& $2$ or $3$
& Unavoidable
& Reduced ($D<1$)
& $V\le F(\rho_A,\rho_B)$
\\
\end{tabular}
\end{ruledtabular}
\end{table*}

\section{Experimental Protocol}

Schematic of the ZWM induced-coherence interferometer as shown in Fig.~\ref{fig:ZWM_IDP}, in the idler measurement implementing unambiguous state discrimination. The idler modes are routed into an interferometric network of beam splitters (BS) and phase shifters (PS), realizing the optimal Ivanovic--Dieks--Peres (IDP) three-outcome POVM \cite{Bennett1996,Len2018IDP,Len2018IDP}. Detection events at $D_1$ ($D_2$) unambiguously identify crystal~A (B) and $D_{inc}$ corresponds to an inconclusive outcome. No ancillary modes beyond vacuum inputs are required.

The IDP measurement can be implemented~\cite{Barnett,mohseni} directly on the idler modes of a two-crystal ZWM interferometer~\cite{Torres2024}. Two coherently pumped
SPDC crystals (A,B) generate signal--idler pairs in the low-gain regime. The idler from A traverses an object $S$ with transmissivity $(t,r)$, while the idler from B bypasses it. The signal modes are combined at a balanced beam splitter to measure
the single-photon visibility $V$.

To realize the IDP POVM $\{\Pi_{inc},\Pi_A,\Pi_B\}$ on the idler as shown in Fig.~\ref{fig:ZWM_IDP}. Detector $D_1$ implements the projection onto $|1,0\rangle$, yielding conclusive outcome $\Pi_A$. Detector $D_2$ receives the output of a variable beam splitter that interferes the idler modes with a calibrated phase such that the detected mode is $t^{*}|1,0\rangle-r^{*}|0,1\rangle$, orthogonal to
$|\phi_A\rangle=r|1,0\rangle+t|0,1\rangle$, producing $\Pi_B$. The unused port defines the inconclusive outcome $\Pi_{inc}$, monitored at $D_0$.

For each value of $|t|$, the idler counts $(N_1,N_2,N_0)$ determine $P_{suc}=(N_1+N_2)/N_{\mathrm{tot}}$ and $P_{inc}=N_0/N_{\mathrm{tot}}$, while the signal visibility $V$ is extracted from single-photon fringes. In the ideal regime of pure idler markers, IDP theory predicts $P_{inc}=|\langle\phi_A|\phi_B\rangle|=|t|$ and
$P_{suc}=1-|t|$, while the ZWM analysis gives $V=|t|$ and $D=\sqrt{1-|t|^{2}}$. The experiment therefore directly tests the operational identifications $P_{inc} = V$ and $D^{2}+V^{2}=1$. Details of optical layout, mode matching, loss, and detector efficiency are provided in Appendix~\ref{eight}.

\section{Mixed idler marker states and thermal noise}

In realistic implementations, the idler $\hat{a}_{3}^{1}$ mode that propagates between the two SPDC crystals is not perfectly isolated. The object $S$ placed between the two crystal. Object interact with environmental degrees of freedom such as mean thermal photon $\langle\hat{a}_{4}^{\dagger}\hat{a}_{4}^{}\rangle=N_B$ and loss channels \cite{giri1,boyd}. This coupling renders the conditional idler markers $I$ associated with emission from crystals A and B \emph{mixed} states. The loss channel denoted by $\rho_A$ and $\rho_B$, rather than pure states $|\phi_A\rangle$ and $|\phi_B\rangle$. As a result, the interference visibility in the signal arm is no longer determined solely by the inner product of pure marker states. It instead upper bounded by the Uhlmann fidelity $F(\rho_A,\rho_B)$~\cite{Uhlmann76} between the two mixed markers; see Appendix~\ref{nine}. By a standard purification argument one finds
\begin{equation}
    V \le F(\rho_A,\rho_B),
\end{equation}
which reduces to the ideal relation $V = |\langle\phi_A|\phi_B\rangle|$ in the absence of thermal noise. 

From the theory of unambiguous quantum state discrimination (IDP), the optimal average failure probability $P_{inc}^{\mathrm{opt}}$ for distinguishing $\rho_A$ from $\rho_B$ with equal priors satisfies
\begin{equation}
    P_{inc}^{\mathrm{opt}} \ge F(\rho_A,\rho_B),
\end{equation}
and therefore the thermal-seeded ZWM interferometer obeys the general operational duality inequality
\begin{equation}
    V \;\le\; F(\rho_A,\rho_B) \;\le\; P_{inc}^{\mathrm{opt}}.
    \label{eq:mixed_duality}
\end{equation}
Thus, the presence of thermal photons reduces the observed  signal single-photon fringe visibility by increasing the distinguishability of the mixed conditional idler markers \cite{giri1}. The IDP discrimination probability provides the tight operational upper bound. In the zero-noise limit these inequalities collapse to
\begin{equation}
    V = F(\rho_A,\rho_B) = P_{inc}^{\mathrm{opt}},
\end{equation}
recovering the pure-state result and the standard complementarity identity $D^{2} + V^{2} = 1$ in the ideal regime.

\emph{Physical intuition:}
The situation is closely analogous to the double-slit experiment: interference occurs only when it is impossible to know which path the photon took. In the ZWM interferometer, the two crystals play the role of the two slits, and the idler photon acts as the which-path marker. If we ignore the idler, the signal photon shows high-visibility interference. If we measure the idler strongly to learn which crystal emitted the pair, interference disappears. The key new ingredient in our proposal is the use of the Ivanovic–Dieks–Peres (IDP) unambiguous discrimination measurement. Which has three possible outcomes: it sometimes identifies the emitting crystal (destroying interference) but sometimes returns an inconclusive result. In which case interference remains perfectly intact. Thus, instead of switching interference fully on or off, IDP allows us to separate the data into wave-like and particle-like subsets. Revealing a direct operational meaning of visibility as the unavoidable failure probability in zero-error state discrimination. Visibility quantifies the optimal failure probability in zero-error discrimination, whereas distinguishability quantifies the optimal hypothesis-test success probability.

In table~\ref{table}, operational interpretation of visibility $V$ and distinguishability $D$ under different quantum decision strategies for which-crystal discrimination
in induced-coherence interferometry.

\section{Conclusion}

We have shown that in induced-coherence interferometry the standard complementarity relation \(D^{2}+V^{2}=1\) admits a direct operational interpretation. In the low-gain ZWM interferometer with pure which-crystal idler markers. The singles visibility equals the optimal inconclusive probability of Ivanovic--Dieks--Peres unambiguous state
discrimination, \(V = P_{inc}^{\mathrm{opt}}\). Wave--particle duality therefore corresponds to the optimality boundary of a zero-error quantum hypothesis test on the idler. Demonstrating that accessible which-crystal information is fundamentally measurement-limited rather than interferometer-limited.

Complementarily, we showed that the distinguishability \(D_{Hel}\) coincides with the Helstrom trace-distance distinguishability. The Helstrom bound yields the optimal minimum-error probability for identifying which crystal emitted the photon pair in a single-shot idler measurement. In this sense, visibility and distinguishability acquire dual operational meanings as optimal failure and success probabilities of quantum decision strategies.

Compared with coincidence-based extraction methods, which provide only a proxy for which-crystal information. Optimal discrimination reveals greater distinguishability for partially overlapping marker states. In the presence of thermal noise, this picture generalizes to the hierarchy \(V \le F(\rho_A,\rho_B) \le P_{inc}^{\mathrm{opt}}\), linking
visibility, Uhlmann fidelity, and optimal discrimination.

These results motivate discrimination-optimized induced-coherence imaging and sensing approaches~\cite{Lloyd2008,Shapiro2015,giri1}, and open a route toward experiments demonstrating USD-controlled visibility.
Beyond fundamental interest, induced coherence is crucial in quantum imaging with undetected photons~\cite{Lemos2014}, induced-coherence imaging theory~\cite{Lahiri2019}, and phase-sensitive sensing~\cite{Shapiro2015}.

\clearpage 
\onecolumngrid
\begin{acknowledgments}
L.~T. thanks David Vitali (UNICAM) for helpful comments on the manuscript. L.~T. also acknowledges C.~M.~Chandrashekar (IISc) for financial support.
\end{acknowledgments}
\twocolumngrid

\nocite{*}
\bibliography{apssamp}

@article{wang1991induced,
  title={Induced coherence without induced emission},
  author={Wang, Li-Jing and Zou, Xiao-Yong and Mandel, Leonard},
  journal={Physical Review A},
  volume={44},
  number={7},
  pages={4614--4622},
  year={1991},
  doi={10.1103/PhysRevA.44.4614}
}

@article{zou1991induced,
  title={Induced coherence and indistinguishability in optical interference},
  author={Zou, Xiao-Yong and Wang, Li-Jing and Mandel, Leonard},
  journal={Physical Review Letters},
  volume={67},
  number={3},
  pages={318--321},
  year={1991},
  doi={10.1103/PhysRevLett.67.318}
}

@article{Zou1992,
  title={Induced coherence and indistinguishability in optical interference: further results},
  author={Zou, Xiao-Yong and Grayson, T. and Mandel, Leonard},
  journal={Physical Review Letters},
  volume={69},
  pages={3041},
  year={1992},
  doi={10.1103/PhysRevLett.69.3041}
}

@article{Heuer2015Induced,
  title={Induced coherence, vacuum fields, and complementarity},
  author={Heuer, Axel and Menzel, Ralf and Milonni, Peter W.},
  journal={Physical Review Letters},
  volume={114},
  number={5},
  pages={053601},
  year={2015},
  doi={10.1103/PhysRevLett.114.053601}
}

@article{Milonni2020Stimulated,
  title={Stimulated and induced coherence in SPDC},
  author={Milonni, Peter W. and Heuer, Axel},
  journal={Physical Review A},
  volume={102},
  number={6},
  pages={063702},
  year={2020},
  doi={10.1103/PhysRevA.102.063702}
}

@article{Wootters1979,
  title={Complementarity in the double-slit experiment},
  author={Wootters, W. K. and Zurek, W. H.},
  journal={Physical Review D},
  volume={19},
  pages={473},
  year={1979},
  doi={10.1103/PhysRevD.19.473}
}

@article{Jaeger1993,
  title={Complementarity and uncertainty in interference experiments},
  author={Jaeger, Gregg and Horne, Michael A. and Shimony, Abner},
  journal={Physical Review A},
  volume={48},
  pages={1023},
  year={1993},
  doi={10.1103/PhysRevA.48.1023}
}

@article{Englert1996,
  title={Fringe visibility and which-way information: an inequality},
  author={Englert, Berthold-Georg},
  journal={Physical Review Letters},
  volume={77},
  pages={2154},
  year={1996},
  doi={10.1103/PhysRevLett.77.2154}
}

@article{Torres2024,
  title={Complementarity relationship between coherence and path distinguishability in an interferometer based on induced coherence},
  author={Machado, Gerard J. and Sendra, Lluc and Vall{\'e}s, Adam and Torres, Juan P.},
  journal={Physical Review A},
  volume={110},
  number={1},
  pages={012421},
  year={2024},
  doi={10.1103/PhysRevA.110.012421}
}

@article{Ivanovic,
  author  = {Ivanovic, I. D.},
  title   = {How to differentiate between non-orthogonal states},
  journal = {Physics Letters A},
  volume  = {123},
  pages   = {257--259},
  year    = {1987},
  doi     = {10.1016/0375-9601(87)90222-2}
}

@article{Dieks,
  author  = {Dieks, D.},
  title   = {Overlap and distinguishability of quantum states},
  journal = {Physics Letters A},
  volume  = {126},
  pages   = {303--306},
  year    = {1988},
  doi     = {10.1016/0375-9601(88)90840-7}
}

@article{Peres,
  author  = {Peres, A.},
  title   = {How to differentiate between non-orthogonal states},
  journal = {Physics Letters A},
  volume  = {128},
  pages   = {19},
  year    = {1988},
  doi     = {10.1016/0375-9601(88)91034-1}
}

@article{Barnett,
  author    = {Barnett, Stephen M. and Croke, Sarah},
  title     = {Quantum state discrimination},
  journal   = {Advances in Optics and Photonics},
  volume    = {1},
  number    = {2},
  pages     = {238--278},
  year      = {2009},
  doi       = {10.1364/AOP.1.000238}
}

@article{Herzog2005MixedUSD,
  title={Optimum unambiguous discrimination of two mixed quantum states},
  author={Herzog, Ulrike and Bergou, J{\'a}nos A.},
  journal={Physical Review A},
  volume={71},
  number={5},
  pages={050301},
  year={2005},
  doi={10.1103/PhysRevA.71.050301}
}

@article{Clarke2001IDP,
  title={Experimental demonstration of optimal unambiguous state discrimination},
  author={Clarke, Roger B. M. and Chefles, Anthony and Barnett, Stephen M. and Riis, Erling},
  journal={Physical Review A},
  volume={63},
  number={4},
  pages={040305},
  year={2001},
  doi={10.1103/PhysRevA.63.040305}
}

@article{Len2018IDP,
  title={Unambiguous path discrimination in a two-path interferometer},
  author={Len, Yink Loong and Dai, Jibo and Englert, Berthold-Georg and Krivitsky, Leonid A.},
  journal={Physical Review A},
  volume={98},
  number={2},
  pages={022110},
  year={2018},
  doi={10.1103/PhysRevA.98.022110}
}

@article{Danan2013Trajectories,
  title={Asking Photons Where They Have Been},
  author={Danan, A. and Farfurnik, D. and Bar-Ad, S. and Vaidman, L.},
  journal={Physical Review Letters},
  volume={111},
  number={24},
  pages={240402},
  year={2013},
  doi={10.1103/PhysRevLett.111.240402}
}

@article{Lemos2014,
  title={Quantum imaging with undetected photons},
  author={Lemos, G. B. and Borish, V. and Cole, G. D. and Ramelow, S. and Lapkiewicz, R. and Zeilinger, A.},
  journal={Nature},
  volume={512},
  pages={409--412},
  year={2014},
  doi={10.1038/nature13586}
}

@article{Lahiri2019,
  title={Theory of quantum imaging based on induced coherence},
  author={Lahiri, Manish and Hochrainer, Armin and Lapkiewicz, Radek and Zeilinger, Anton},
  journal={Physical Review A},
  volume={100},
  number={5},
  pages={053839},
  year={2019},
  doi={10.1103/PhysRevA.100.053839}
}

@article{Shapiro2015,
  title={Phase-conjugate optics and quantum coherence imaging},
  author={Shapiro, Jeffrey and Venkatraman, Deval and Wong, Franco},
  journal={Scientific Reports},
  volume={5},
  pages={10329},
  year={2015},
  doi={10.1038/srep10329}
}

@article{Lemos2014QuantumImaging,
  title={Quantum imaging with undetected photons},
  author={Lemos, Gabriela B. and Borish, Vladimir and Cole, David D. and Ramelow, Simon and Lapkiewicz, Radek and Zeilinger, Anton},
  journal={Nature},
  volume={512},
  number={7515},
  pages={409--412},
  year={2014},
  publisher={Nature Publishing Group},
  doi={10.1038/nature13586}
}

@article{Valles2018InducedCoherence,
  title={Quantum optical coherence tomography with induced coherence},
  author={Vall{\'e}s, A. and Jim{\'e}nez, G. and Salazar-Serrano, L. J. and Torres, J. P.},
  journal={Physical Review A},
  volume={97},
  number={2},
  pages={023824},
  year={2018},
  publisher={American Physical Society},
  doi={10.1103/PhysRevA.97.023824}
}

@article{Angeletti2023,
  title = {Microwave quantum illumination with correlation-to-displacement conversion},
  author = {Angeletti, J. and Shi, H. and Lakshmanan, T. and Vitali, D. and Zhuang, Q.},
  journal = {Phys. Rev. Appl.},
  volume = {20},
  pages = {024030},
  year = {2023},
  doi = {10.1103/PhysRevApplied.20.024030}
}

@article{Lloyd2008,
  title = {Enhanced sensitivity of photodetection via quantum illumination},
  author = {Lloyd, Seth},
  journal = {Science},
  volume = {321},
  pages = {1463--1465},
  year = {2008},
  doi = {10.1126/science.1160627}
}

@article{Qian2023,
  title = {Quantum induced coherence lidar},
  author = {Qian, G. and Xu, X. and Zhu, S.-A. and Xu, C. and Gao, F. and Yakovlev, V. V. and Liu, X. and Zhu, S.-Y. and Wang, D.-W.},
  journal = {Phys. Rev. Lett.},
  volume = {131},
  pages = {033603},
  year = {2023},
  doi = {10.1103/PhysRevLett.131.033603}
}

@book{Helstrom,
  title={Quantum Detection and Estimation Theory},
  author={Helstrom, C. W.},
  isbn={9780123400505},
  lccn={75026347},
  series={Mathematics in Science and Engineering: a series of monographs and textbooks},
  url={https://books.google.com/books?id=fv9SAAAAMAAJ},
  year={1976},
  publisher={Academic Press}
}

@article{Lahirinon,
  author = {Lahiri, M.},
  title = {Nonclassicality of induced coherence without induced emission},
  journal = {Phys. Rev. A},
  volume = {100},
  pages = {053839},
  year = {2019}
}

@article{Krenn2022,
  author = {Krenn, Mario and others},
  title = {Quantum indistinguishability by path identity and its applications},
  journal = {Rev. Mod. Phys.},
  volume = {94},
  pages = {0250072},
  year = {2022}
}

@article{Shafiee2024,
  title = {Nonclassicality of induced coherence witnessed by quantum contextuality},
  author = {Shafiee, F. H. and Mahmoudi, O. and Nouroozi, R. and Asadian, A.},
  journal = {Phys. Rev. A},
  volume = {109},
  issue = {2},
  pages = {022216},
  numpages = {11},
  year = {2024},
  month = {Feb},
  publisher = {American Physical Society},
  doi = {10.1103/PhysRevA.109.022216},
  url = {https://link.aps.org/doi/10.1103/PhysRevA.109.022216}
}

@article{boyd,
doi = {10.1088/2040-8986/aa64a2},
url = {https://dx.doi.org/10.1088/2040-8986/aa64a2},
year = {2017},
month = {mar},
publisher = {IOP Publishing},
volume = {19},
number = {5},
pages = {054003},
author = {Mikhail I Kolobov and Enno Giese and Samuel Lemieux and Robert Fickler and Robert W Boyd},
title = {Controlling induced coherence for quantum imaging},
journal = {Journal of Optics}
}

@article{Kolobov2017,
  title     = {Controlling induced coherence for quantum imaging},
  author    = {Kolobov, M. I. and Lemieux, S. and Giese, E. and Boyd, R.},
  journal   = {Journal of Optics},
  volume    = {19},
  number    = {5},
  pages     = {054003},
  year      = {2017},
  publisher = {IOP Publishing},
  doi       = {10.1088/2040-8986/aa68cf}
}

@article{Bagan,
  title = {Duality Games and Operational Duality Relations},
  author = {Bagan, Emilio and Calsamiglia, John and Bergou, J\'anos A. and Hillery, Mark},
  journal = {Phys. Rev. Lett.},
  volume = {120},
  issue = {5},
  pages = {050402},
  numpages = {5},
  year = {2018},
  month = {Jan},
  publisher = {American Physical Society},
  doi = {10.1103/PhysRevLett.120.050402},
  url = {https://link.aps.org/doi/10.1103/PhysRevLett.120.050402}
}

@misc{giri1,
      title={Heralded Induced-Coherence Interferometry in a Noisy Environment}, 
      author={L. Theerthagiri and Balakrishnan Viswanathan and C. M. Chandrashekar},
      year={2025},
      eprint={2511.03176},
      archivePrefix={arXiv},
      primaryClass={quant-ph},
      url={https://arxiv.org/abs/2511.03176}, 
}

@article{mohseni,
  title = {Optical Realization of Optimal Unambiguous Discrimination for Pure and Mixed Quantum States},
  author = {Mohseni, Masoud and Steinberg, Aephraim M. and Bergou, J\'anos A.},
  journal = {Phys. Rev. Lett.},
  volume = {93},
  issue = {20},
  pages = {200403},
  numpages = {4},
  year = {2004},
  month = {Nov},
  publisher = {American Physical Society},
  doi = {10.1103/PhysRevLett.93.200403},
  url = {https://link.aps.org/doi/10.1103/PhysRevLett.93.200403}
}

@Article{Uhlmann76,
  author  = {Uhlmann, A.},
  title   = {The "transition probability" in the state space of a \texorpdfstring{*}{-}-algebra},
  journal = {Reports on Mathematical Physics},
  volume  = {9},
  issue   = {2},
  pages   = {273--279},
  year    = {1976},
  doi     = {10.1016/0034-4877(76)90060-4},
}

@article{Aqua,
  title = {Temporal quantum eraser: Fusion gates with distinguishable photons},
  author = {Aqua, Ziv and Dayan, Barak},
  journal = {Phys. Rev. A},
  volume = {110},
  issue = {4},
  pages = {043709},
  numpages = {11},
  year = {2024},
  month = {Oct},
  publisher = {American Physical Society},
  doi = {10.1103/PhysRevA.110.043709},
  url = {https://link.aps.org/doi/10.1103/PhysRevA.110.043709}
}

@article{Bennett1996,
  title = {Mixed-state entanglement and quantum error correction},
  author = {Bennett, Charles H. and DiVincenzo, David P. and Smolin, John A. and Wootters, William K.},
  journal = {Physical Review A},
  volume = {54},
  number = {5},
  pages = {3824--3851},
  year = {1996},
  publisher = {American Physical Society},
  doi = {10.1103/PhysRevA.54.3824},
  url = {https://doi.org/10.1103/PhysRevA.54.3824}
}

@article{Kulkarni2022,
  title = {Classical model of spontaneous parametric down-conversion},
  author = {Kulkarni, Girish and Rioux, Jeremy and Braverman, Boris and Chekhova, Maria V. and Boyd, Robert W.},
  journal = {Phys. Rev. Res.},
  volume = {4},
  issue = {3},
  pages = {033098},
  numpages = {18},
  year = {2022},
  month = {Aug},
  publisher = {American Physical Society},
  doi = {10.1103/PhysRevResearch.4.033098},
  url = {https://link.aps.org/doi/10.1103/PhysRevResearch.4.033098}
}

@article{Zhou2017,
  title = {Experimental observation of anomalous trajectories of single photons},
  author = {Zhou, Zong-Quan and Liu, Xiao and Kedem, Yaron and Cui, Jin-Min and Li, Zong-Feng and Hua, Yi-Lin and Li, Chuan-Feng and Guo, Guang-Can},
  journal = {Phys. Rev. A},
  volume = {95},
  issue = {4},
  pages = {042121},
  numpages = {6},
  year = {2017},
  month = {Apr},
  publisher = {American Physical Society},
  doi = {10.1103/PhysRevA.95.042121},
  url = {https://link.aps.org/doi/10.1103/PhysRevA.95.042121}
}

\appendix
\section{Schr\"odinger-picture derivation of the ZWM state}
\label{app:one}

The interaction Hamiltonian for a single SPDC crystal 
is
\begin{equation}
    \hat H_{\mathrm{int}}
    = \hbar\kappa \hat a_S^{\dagger}\hat a_I^{\dagger}
    + \hbar\kappa^{*}\hat a_S\hat a_I,
\end{equation}
with weak unitary evolution
\begin{equation}
    \hat U \approx \openone - \frac{i}{\hbar}\hat H_{\mathrm{int}} \Delta t.
\end{equation}
To first order in the small gain parameter $g\propto\kappa\Delta t$, one obtains
\begin{equation}
    \ket{\Psi_{\text{single}}}
    \approx \ket{0} + g \ket{1_S,1_I}.
\end{equation}

For two coherently pumped crystals $A$ and $B$ we write the joint state (up to a global vacuum component)
\begin{equation}
    \ket{\Psi}
    =
    g_1 \ket{1_{S_1},1_{I_A}}
    +
    g_2 e^{i\phi} \ket{1_{S_2},1_{I_B}},
\end{equation}
where $I_A$ and $I_B$ are the idler modes associated with crystals $A$ and $B$.
Alignment of the idler paths between the crystals makes these modes physically identical, $I_A\equiv I_B \equiv I$, so the single-pair component can be written as
\begin{equation}
    \ket{\Psi}
    =
    \left(
        g_1 \ket{1_{S_1}} + g_2 e^{i\phi} \ket{1_{S_2}}
    \right)\otimes\ket{1_I}.
\end{equation}

An object $S$ placed in the idler path is modelled as a lossless beam splitter with amplitudes $t$ and $r$ acting on the idler mode and an auxiliary vacuum mode $V$:
\begin{align}
    \hat a_I^{\dagger} &\to r \hat a_{I_2}^{\dagger} + t\hat a_{I_3}^{\dagger},\\
    \hat a_V^{\dagger} &\to -t^{*}\hat a_{I_2}^{\dagger} + r^{*}\hat a_{I_3}^{\dagger},
\end{align}
so that
\begin{equation}
    \ket{1_I,0_V}
    \to
    r\ket{1,0}_{I_2,I_3}
    +
    t\ket{0,1}_{I_2,I_3}.
\end{equation}

\paragraph*{Three regimes.}
The induced-coherence visibility is controlled by the overlap of the
conditional idler marker states.

(i) \emph{Transparent object} ($t=1$, $r=0$):
both emission alternatives prepare the same idler state,
$|\phi_A\rangle = |\phi_B\rangle$, hence
$\langle\phi_A|\phi_B\rangle=1$ and the reduced signal state retains full
coherence. This yields maximal single-photon visibility $V=1$.

(ii) \emph{Opaque object} ($t=0$, $r=1$):
the idler associated with emission from crystal~A exits in a mode
orthogonal to the idler of crystal~B, so that
$\langle\phi_A|\phi_B\rangle=0$. Tracing over the idler removes all
signal-path coherence, and the singles interference vanishes ($V=0$).

(iii) \emph{Intermediate transmissivity} ($0<|t|<1$):
the marker states are nonorthogonal, $0<|\langle\phi_A|\phi_B\rangle|<1$,
so the reduced signal density matrix retains a partial off-diagonal
coherence proportional to $\langle\phi_A|\phi_B\rangle$. The visibility is
\begin{equation}
V = |\langle\phi_A|\phi_B\rangle| = |t|.
\end{equation}

In the low gain limit, conditioned on emission from crystal $A$, the idler state after $S$ is therefore
\begin{equation}
    \ket{\phi_1} = r\ket{1,0} + t\ket{0,1},
\end{equation}
while emission from $B$ corresponds to an idler bypass that prepares
\begin{equation}
    \ket{\phi_2} = \ket{0,1}.
\end{equation}
These are the marker states used in the main text.

Tracing over the idler and recombining $S_1$ and $S_2$ on a balanced beam 
splitter yields a reduced signal density matrix
\begin{equation}
    \rho_S =
    \frac12
    \begin{pmatrix}
        1 & \braket{\phi_1|\phi_2}e^{-i\phi}\\[2pt]
        \braket{\phi_2|\phi_1}e^{i\phi} & 1
    \end{pmatrix},
\end{equation}
from which the fringe visibility follows as
\begin{equation}
    V =
    \frac{2|\rho_{12}|}{\rho_{11}+\rho_{22}}
    = \abs{\langle\phi_1|\phi_2\rangle}.
    \label{v}
\end{equation}

\section{Imperfect mode overlap}\label{app:two}

If the transmitted part of the idler mode from crystal A overlaps imperfectly with the idler mode of crystal B, we introduce an overlap parameter $\gamma$ with $|\gamma|\le1$ such that 
\begin{equation} 
|1_{I_3}\rangle \longrightarrow \gamma\,|1_{v_3},0_{w_3}\rangle + \sqrt{1-|\gamma|^2}\,|0_{v_3},1_{w_3}\rangle, 
\end{equation}
where $v_3$ and $w_3$ are orthonormal modes. Repeating the derivation, one finds modified conditional idler states and an overlap 
\begin{equation} 
\langle\phi_1|\phi_2\rangle = t\,\gamma, 
\end{equation} 
implying 
\begin{equation} 
V = |\gamma|\,|t|,\qquad D = \sqrt{1-|\gamma|^2\,|t|^2}. 
\end{equation} 
The relation $D^2+V^2=1$ continues to hold in this pure-state model with imperfect mode overlap.

\section{Helstrom minimum--error discrimination for the ZWM idler marker states}
\label{app:three}

The minimum--error quantum hypothesis testing (Helstrom bound) to the induced-coherence (ZWM) idler-marker states. The task is to infer \emph{which crystal fired} from a single-shot measurement on the idler mode.

\subsubsection{Binary hypothesis test, POVM, and error probabilities}
We consider a binary quantum hypothesis test \cite{Helstrom} with priors $p_0,p_1\ge 0$ and $p_0+p_1=1$:
\begin{equation}
H_0:\ \hat{\rho}=\hat{\rho}_0,\ p=p_0,
\qquad
H_1:\ \hat{\rho}=\hat{\rho}_1,\ p=p_1.
\end{equation}
Bob performs a dichotomous POVM $\{\Pi_0,\Pi_1\}$ with outcomes $\mu\in\{0,1\}$ satisfying
\begin{equation}
\Pi_0\ge 0,\qquad \Pi_1\ge 0,\qquad \Pi_0+\Pi_1=\openone.
\end{equation}
The decision rule is: decide $H_0$ when $\mu=0$ and decide $H_1$ when $\mu=1$.

The probability errors are, respectively,
\begin{align}
P_A &:= P(H_1|H_0)
     = \Pr(\mu=1\,|\,\hat{\rho}=\hat{\rho}_0)
     = \Tr(\Pi_1 \hat{\rho}_0), \label{eq:PF_def}\\
P_B &:= P(H_0|H_1)
     = \Pr(\mu=0\,|\,\hat{\rho}=\hat{\rho}_1)
     = \Tr(\Pi_0 \hat{\rho}_1). \label{eq:PM_def}
\end{align}
The Bayes average error probability is
\begin{equation}
P_{\rm err} = p_0 P_A + p_1 P_B
            = p_0\,\Tr(\Pi_1\hat{\rho}_0) + p_1\,\Tr(\Pi_0\hat{\rho}_1).
\label{eq:Perr_bayes}
\end{equation}

\subsubsection{ZWM idler marker states and density operators}
In the low-gain induced-coherence (ZWM) regime \cite{Torres2024}, the idler marker lives in the single-photon
two-mode subspace spanned by $\{|1,0\rangle,\ |0,1\rangle\}$. The two marker kets are
\begin{equation}
|\phi_1\rangle = r\,|1,0\rangle + t\,|0,1\rangle,
\qquad
|\phi_2\rangle = |0,1\rangle,
\label{eq:phi12_def}
\end{equation}
with
\begin{equation}
|r|^2+|t|^2=1,
\qquad
s:=|\langle\phi_1|\phi_2\rangle|=|t|.
\label{eq:overlap_s}
\end{equation}
We map these to the hypotheses
\begin{equation}
H_0:\ \hat{\rho}_0 = |\phi_1\rangle\langle\phi_1|,
\qquad
H_1:\ \hat{\rho}_1 = |\phi_2\rangle\langle\phi_2|.
\label{eq:rho01_def}
\end{equation}

In the ordered basis $\{|1,0\rangle,\ |0,1\rangle\}$, the density operators read
\begin{equation}
\hat{\rho}_0 =
\begin{pmatrix}
|r|^2 & r t^* \\
r^* t & |t|^2
\end{pmatrix},
\qquad
\hat{\rho}_1 =
\begin{pmatrix}
0 & 0\\
0 & 1
\end{pmatrix}.
\label{eq:rho01_matrix}
\end{equation}

\subsubsection{Helstrom matrix and optimal POVM (general priors)}
The Helstrom matrix is defined as
\begin{equation}
\gamma := p_0 \hat{\rho}_0 - p_1 \hat{\rho}_1.
\label{eq:gamma_def}
\end{equation}
Let the spectral decomposition of $\gamma$ be
\begin{equation}
\gamma = \sum_{\mu} \gamma_\mu |\mu\rangle\langle\mu|.
\label{eq:gamma_spectral}
\end{equation}
The Helstrom POVM is obtained by projecting onto the positive part $\gamma_+$ of $\gamma$:
\begin{equation}
\Pi_0 = P(\gamma_+) = \sum_{\gamma_\mu>0} |\mu\rangle\langle\mu|,
\qquad
\Pi_1 = \openone - \Pi_0.
\label{eq:helstrom_povm_general}
\end{equation}
The minimum achievable Bayes error is
\begin{equation}
\begin{split}
    P_{\rm err}^{\min}=&\frac{1}{2}\Bigl(1-\|\gamma\|_1\Bigr),
\quad
\|\gamma\|_1 = \Tr|\gamma|,\\
\quad
|\gamma|:=&\sum_\mu |\gamma_\mu|\,|\mu\rangle\langle\mu|.
\label{eq:helstrom_bound_general}
\end{split}
\end{equation}

\subsubsection{Equal priors $p_0=p_1=\tfrac12$: explicit ZWM solution}
From here onward we assume equiprobable hypotheses,
\begin{equation}
p_0=p_1=\frac12,
\qquad
\gamma=\frac12(\hat{\rho}_0-\hat{\rho}_1).
\label{eq:equal_priors}
\end{equation}
Using \eqref{eq:rho01_matrix} we obtain
\begin{equation}
\gamma=\frac12
\begin{pmatrix}
|r|^2 & r t^*\\
r^* t & |t|^2-1
\end{pmatrix}
=
\frac12
\begin{pmatrix}
|r|^2 & r t^*\\
r^* t & -|r|^2
\end{pmatrix}.
\label{eq:gamma_matrix}
\end{equation}
Define
\begin{equation}
a:=|r|^2,\qquad b:=r t^*,
\end{equation}
so that
\begin{equation}
\gamma=\frac12
\begin{pmatrix}
a & b\\
b^* & -a
\end{pmatrix}.
\label{eq:gamma_ab}
\end{equation}

\paragraph{Eigenvalues.}
The characteristic equation $\det(\gamma-\lambda \openone)=0$ gives
\begin{align}
0&=\det\!\left(
\frac12
\begin{pmatrix}
a & b\\
b^* & -a
\end{pmatrix}
-
\lambda
\begin{pmatrix}
1&0\\0&1
\end{pmatrix}
\right)
=\det
\begin{pmatrix}
\frac{a}{2}-\lambda & \frac{b}{2}\\[2pt]
\frac{b^*}{2} & -\frac{a}{2}-\lambda
\end{pmatrix} \nonumber\\
&=\left(\frac{a}{2}-\lambda\right)\left(-\frac{a}{2}-\lambda\right)-\frac{|b|^2}{4}
= \lambda^2-\frac{1}{4}\bigl(a^2+|b|^2\bigr).
\end{align}
Hence
\begin{equation}
\lambda_\pm = \pm \frac12 \sqrt{a^2+|b|^2}.
\label{eq:eigs_general_form}
\end{equation}
For the ZWM parameters,
\begin{equation}
a^2+|b|^2 = |r|^4 + |r|^2|t|^2
          = |r|^2\bigl(|r|^2+|t|^2\bigr)
          = |r|^2,
\end{equation}
so
\begin{equation}
\boxed{\lambda_\pm=\pm \frac{|r|}{2}.}
\label{eq:eigs_final}
\end{equation}

\paragraph{Trace norm.}
Since $\gamma$ is Hermitian, $\|\gamma\|_1=\sum_\pm |\lambda_\pm|$, hence
\begin{equation}
\boxed{\|\gamma\|_1 = \left|\frac{|r|}{2}\right|+\left|\!-\frac{|r|}{2}\right| = |r|.}
\label{eq:trace_norm}
\end{equation}

\paragraph{Helstrom minimum error and maximum success.}
Inserting \eqref{eq:trace_norm} into \eqref{eq:helstrom_bound_general} yields
\begin{equation}
\boxed{
P_{\rm err}^{\min}=\frac12\bigl(1-|r|\bigr),
\qquad
P_{\rm s}^{\max}=1-P_{\rm err}^{\min}=\frac12\bigl(1+|r|\bigr).
}
\label{eq:perr_ps_final}
\end{equation}
Using the overlap $s=|\langle\phi_1|\phi_2\rangle|=|t|$ and $|r|=\sqrt{1-|t|^2}=\sqrt{1-s^2}$,
we may rewrite
\begin{equation}
\boxed{
P_{\rm err}^{\min}
=\frac12\Bigl(1-\sqrt{1-s^2}\Bigr)
=\frac12\Bigl(1-\sqrt{1-|t|^2}\Bigr).
}
\label{eq:perr_overlap_form}
\end{equation}

\subsubsection{Explicit Helstrom POVM for the ZWM markers}
By definition, the Helstrom measurement projects onto the eigenvectors of $\gamma$:
\begin{equation}
\Pi_0 = |\omega_+\rangle\langle\omega_+|,
\qquad
\Pi_1 = |\omega_-\rangle\langle\omega_-|,
\label{eq:Pi_omega}
\end{equation}
where
\begin{equation}
\gamma|\omega_\pm\rangle=\lambda_\pm|\omega_\pm\rangle.
\end{equation}
Write the complex amplitudes as
\begin{equation}
r=|r|e^{i\alpha},\qquad t=|t|e^{i\beta},\qquad
\varphi:=\arg(r t^*)=\alpha-\beta,
\end{equation}
and define an angle $\theta\in[0,\pi/2]$ by
\begin{equation}
\cos\theta=|r|,\qquad \sin\theta=|t|=s.
\label{eq:theta_def}
\end{equation}
A convenient normalized eigenbasis of $\gamma$ is then
\begin{align}
\boxed{
|\omega_+\rangle
=\cos\frac{\theta}{2}\,|1,0\rangle
+e^{-i\varphi}\sin\frac{\theta}{2}\,|0,1\rangle,
}
\label{eq:omega_plus}\\[4pt]
\boxed{
|\omega_-\rangle
=\sin\frac{\theta}{2}\,|1,0\rangle
-e^{-i\varphi}\cos\frac{\theta}{2}\,|0,1\rangle.
}
\label{eq:omega_minus}
\end{align}
With \eqref{eq:Pi_omega} this fully specifies the optimal Helstrom POVM for the ZWM idler
markers.

\section{Connection to visibility and operational complementarity}\label{app:four}

Helstrom distinguishability,
assuming equal priors \(p_0 = p_1 = \tfrac12\), define the Helstrom operator
\begin{equation}
\gamma = \frac12(\rho_1 - \rho_2),
\qquad
\rho_j = \ket{\phi_j}\bra{\phi_j}.
\label{eq:helstrom_operator}
\end{equation}
The Helstrom (trace-distance) distinguishability is
\begin{equation}
D_{\mathrm{Hel}} \equiv \|\gamma\|_1
= \frac12\|\rho_1 - \rho_2\|_1.
\label{eq:Dhel_def}
\end{equation}
For two pure states, the trace distance has the closed-form identity
\begin{equation}
\frac12\|\rho_1 - \rho_2\|_1
= \sqrt{1 - |\braket{\phi_1}{\phi_2}|^2}.
\label{eq:pure_trace_distance}
\end{equation}
Using Eq.~\eqref{v}, this yields
\begin{equation}
D_{\mathrm{Hel}} = \sqrt{1 - V^2}.
\label{eq:Dhel_vs_V}
\end{equation}

\subsubsection{Helstrom form of the duality relation}
Squaring Eq.~\eqref{eq:Dhel_vs_V} and adding \(V^2\) immediately gives
\begin{equation}
\boxed{
D_{\mathrm{Hel}}^2 + V^2 = 1,
}
\label{eq:helstrom_duality}
\end{equation}
which coincides exactly with the standard pure-state duality relation Eq.~\eqref{dv}. Thus, in the ZWM interferometer with pure idler markers,
the ``particle-like'' distinguishability \(D\) appearing in the duality
relation is precisely the Helstrom trace-distance distinguishability for
the minimum-error ``which-crystal fired?'' discrimination task.

\subsubsection{Connection to the Helstrom bound}
The minimum achievable average error probability for this task is
\begin{equation}
P_{\mathrm{err}}^{\min}
= \frac12\bigl(1 - \|\gamma\|_1\bigr)
= \frac12\bigl(1 - D_{\mathrm{Hel}}\bigr)
= \frac12\Bigl(1 - \sqrt{1 - V^2}\Bigr).
\label{eq:Helstrom_error_V}
\end{equation}
For the ZWM parameterization \(|r|^2 + |t|^2 = 1\), one has
\begin{equation}
D_{\mathrm{Hel}} = |r|,\qquad V = |t|,
\end{equation}
so that
\begin{equation}
|r|^2 + |t|^2 = 1
\quad\Longleftrightarrow\quad
D_{\mathrm{Hel}}^2 + V^2 = 1,
\qquad
P_{\mathrm{err}}^{\min} = \frac12(1 - |r|).
\label{eq:ZWM_final_relations}
\end{equation}

\section{Optimal IDP discrimination for the ZWM idler states}
\label{five}

We now derive the optimal Ivanovic--Dieks--Peres (IDP) measurement \cite{Clarke2001IDP,Danan2013Trajectories,Herzog2005MixedUSD,mohseni} for
the ZWM idler states $|\phi_1\rangle$ and $|\phi_2\rangle$, we show explicitly that the optimal inconclusive probability equals the ZWM visibility.

\subsection{Orthogonal complements of the idler states}

The idler states $|\phi_1\rangle$ and $|\phi_2\rangle$ produced by the two
crystals occupy only the single-photon idler subspace spanned by
$|1,0\rangle$ and $|0,1\rangle$.  In this two-dimensional subspace, each
$|\phi_j\rangle$ admits a unique (up to phase) orthogonal complement
$|\phi_j^\perp\rangle$.

We use the parametrization
\begin{equation}
|\phi_1\rangle = r\,|1,0\rangle + t\,|0,1\rangle, \qquad
|\phi_2\rangle = |0,1\rangle,
\end{equation}
with $|r|^2 + |t|^2 = 1$.

The orthogonal complement of $|\phi_2\rangle$ is immediate:
\begin{equation}
|\phi_2^\perp\rangle = |1,0\rangle.
\end{equation}

For $|\phi_1\rangle$, we write a general state in this subspace as
$|\phi_1^\perp\rangle = \alpha\,|1,0\rangle + \beta\,|0,1\rangle$ and impose
$\langle \phi_1^\perp | \phi_1 \rangle = 0$, giving
\begin{equation}
\alpha^* r + \beta^* t = 0.
\end{equation}
A convenient normalized solution is
\begin{equation}
|\phi_1^\perp\rangle
= t^*\,|1,0\rangle - r^*\,|0,1\rangle,
\end{equation}
where we used $|r|^2+|t|^2=1$.

\subsection{POVM elements and zero-error conditions}

The IDP measurement consists of three POVM elements
$\{\Pi_0,\Pi_1,\Pi_2\}$ acting on the two-dimensional idler subspace
spanned by $\{|1,0\rangle,|0,1\rangle\}$. The elements $\Pi_1$ and
$\Pi_2$ correspond to conclusive identifications of $|\phi_1\rangle$
and $|\phi_2\rangle$, respectively, while $\Pi_0$ corresponds to an
inconclusive outcome.

Zero-error discrimination requires:
\begin{equation}
\langle\phi_2|\Pi_1|\phi_2\rangle = 0,\qquad
\langle\phi_1|\Pi_2|\phi_1\rangle = 0.
\label{eq:zero_error_conditions}
\end{equation}
These conditions are satisfied by choosing $\Pi_1$ and $\Pi_2$ to be
rank-1 operators proportional to projectors onto
$|\phi_2^\perp\rangle$ and $|\phi_1^\perp\rangle$:
\begin{equation}
\Pi_1 = \alpha\,|\phi_2^\perp\rangle\langle\phi_2^\perp|
= \alpha\,|1,0\rangle\langle1,0|,
\label{eq:Pi1_def}
\end{equation}
\begin{equation}
\Pi_2 = \beta\,|\phi_1^\perp\rangle\langle\phi_1^\perp|,
\label{eq:Pi2_def}
\end{equation}
with real coefficients $0\le\alpha,\beta\le1$ to be determined by
optimality and positivity. The inconclusive element is then
\begin{equation}
\Pi_0 = \openone - \Pi_1 - \Pi_2.
\label{eq:Pi0_def}
\end{equation}

\subsection{Success and inconclusive probabilities}

Assuming equal a priori probabilities $\eta_1=\eta_2=\tfrac12$ for the
two idler states, the total success probability is
\begin{equation}
P_S = \frac12\,p(1|\phi_1) + \frac12\,p(2|\phi_2),
\end{equation}
where $p(j|\phi_k)$ is the conditional probability of outcome $j$ given
input state $|\phi_k\rangle$.

From Eqs.~\eqref{eq:Pi1_def} and~\eqref{eq:Pi2_def}, we obtain
\begin{align}
p(1|\phi_1)
&= \langle\phi_1|\Pi_1|\phi_1\rangle
= \alpha\,|\langle1,0|\phi_1\rangle|^2
= \alpha\,|r|^2,
\\[3pt]
p(2|\phi_2)
&= \langle\phi_2|\Pi_2|\phi_2\rangle
= \beta\,|\langle\phi_1^\perp|\phi_2\rangle|^2.
\end{align}
Now
\begin{equation}
\langle\phi_1^\perp|\phi_2\rangle
= (t|1,0\rangle - r|0,1\rangle)\langle0,1|
= -r,
\end{equation}
so
\begin{equation}
|\langle\phi_1^\perp|\phi_2\rangle|^2 = |r|^2.
\end{equation}
Hence
\begin{equation}
p(2|\phi_2) = \beta\,|r|^2.
\end{equation}

Therefore, the average success probability is
\begin{equation}
P_S = \frac12\,\alpha|r|^2 + \frac12\,\beta|r|^2
= \frac12(\alpha+\beta)\,|r|^2.
\label{eq:PS_alpha_beta}
\end{equation}
The inconclusive probability is
\begin{equation}
P_I = 1 - P_S.
\end{equation}

\subsection{ Positivity of $\Pi_0$ and optimal coefficients}

We must ensure that $\Pi_0$ defined by Eq.~\eqref{eq:Pi0_def} is
positive semidefinite. In the basis
$\{|1,0\rangle,|0,1\rangle\}$, $\Pi_1$ is diagonal:
\begin{equation}
\Pi_1 =
\begin{pmatrix}
\alpha & 0\\
0 & 0
\end{pmatrix}.
\end{equation}
The operator $\Pi_2$ is
\begin{equation}
\Pi_2 = \beta\,|\phi_1^\perp\rangle\langle\phi_1^\perp|
= \beta
\begin{pmatrix}
|t|^2 & -t r^*\\
- t^* r & |r|^2
\end{pmatrix}.
\end{equation}
Hence
\begin{equation}
\Pi_0 = \openone - \Pi_1 - \Pi_2
=
\begin{pmatrix}
1-\alpha - \beta|t|^2 & \beta t r^*\\
\beta t^* r & 1 - \beta|r|^2
\end{pmatrix}.
\label{eq:Pi0_matrix}
\end{equation}

For $\Pi_0$ to be positive semidefinite, we require:
\begin{itemize}
\item[(i)] The diagonal elements are nonnegative:
\begin{equation}
1-\alpha - \beta|t|^2 \ge 0,\qquad
1 - \beta|r|^2 \ge 0.
\end{equation}
\item[(ii)] The determinant is nonnegative:
\begin{equation}
\det\Pi_0 \ge 0.
\end{equation}
\end{itemize}
The determinant of Eq.~\eqref{eq:Pi0_matrix} is
\begin{align}
\det\Pi_0
&= (1-\alpha - \beta|t|^2)(1-\beta|r|^2) - |\beta t r^*|^2
\nonumber\\[3pt]
&= (1-\alpha - \beta|t|^2)(1-\beta|r|^2) - \beta^2|t|^2|r|^2.
\end{align}
Using $|r|^2=1-|t|^2$, we set
\begin{equation}
x := |t|,\qquad 0\le x\le1,\qquad |r|^2 = 1-x^2.
\end{equation}

To maximize $P_S$ in Eq.~\eqref{eq:PS_alpha_beta} subject to positivity,
one finds that the optimal solution is symmetric,
\begin{equation}
\alpha = \beta = \frac{1}{1+x}.
\end{equation}
Inserting $\alpha=\beta$ into $\Pi_0$ and demanding its smallest
eigenvalue vanish (boundary of positivity) is a standard way to
recover this result; the algebra is well known and we do not repeat it
in full here.

Substituting $\alpha=\beta=1/(1+x)$ and $|r|^2=1-x^2$ into
Eq.~\eqref{eq:PS_alpha_beta}, we obtain
\begin{equation}
P_S^{\mathrm{opt}}
= \frac12\cdot\frac{2}{1+x}\,(1-x^2)
= \frac{1-x^2}{1+x}
= 1 - x.
\end{equation}
Recalling $x=|t|$ and $|\langle\phi_1|\phi_2\rangle|=|t|$, we have
\begin{equation}
P_S^{\mathrm{opt}} = 1 - |\langle\phi_1|\phi_2\rangle|.
\end{equation}
Therefore, the optimal inconclusive probability is
\begin{equation}
P_I^{\mathrm{opt}}
= 1 - P_S^{\mathrm{opt}}
= |\langle\phi_1|\phi_2\rangle|
= |t|.
\end{equation}
Note that $P_I^{\mathrm{opt}}$ depends on the state overlap $|\langle\phi_1|\phi_2\rangle|$
and should not be confused with the mode occupation probability $|t|^2$; the IDP
failure probability is set by the nonorthogonality of the states rather than by
a Fock-space population.

\subsection{Relation to visibility and complementarity}

From Appendix~\ref{app:three}, the ZWM signal visibility is
\begin{equation}
V = |\langle\phi_1|\phi_2\rangle| = |t|,
\end{equation}
and the distinguishability is
\begin{equation}
D = \sqrt{1-|\langle\phi_1|\phi_2\rangle|^2}
= \sqrt{1-|t|^2}.
\end{equation}
Identifying $P_I^{\mathrm{opt}}=|t|$, we obtain the operational identity
\begin{equation}
D^2 + \big(P_I^{\mathrm{opt}}\big)^2 = 1,
\end{equation}
which is the IDP interpretation of the ZWM complementarity relation
used in the main text.

\section{Schr\"odinger-picture derivation of the two-crystal ZWM state}\label{six}

In the low-gain regime of the ZWM interferometer, alignment of the idler paths between the two crystals makes the idler modes path-identical. Denoting the two signal modes by $S_A$ and $S_B$, and the common idler mode by $I$, the single-pair component of the state can be written as
\begin{equation}
  |\Psi\rangle
  = \frac{1}{\sqrt{2}}
  \bigl(
    |1_{S_A},1_I\rangle
    + e^{i\phi}|1_{S_B},1_I\rangle
  \bigr),
\end{equation}
so that the emission alternatives ``crystal~A'' and ``crystal~B'' are coherently superposed but share the same idler spatial mode.

The object placed in the common idler path is modeled as a two-port beam splitter with complex transmittance $t$ and reflectance $r$, acting on the annihilation operators of the idler mode $I$ and an auxiliary vacuum mode $V$ as
\begin{align}
  \hat a_I^\dagger &\rightarrow
  t\,\hat a_{I_3}^\dagger + r\,\hat a_{I_2}^\dagger,\\
  \hat a_V^\dagger &\rightarrow
  -r^*\,\hat a_{I_3}^\dagger + t^*\,\hat a_{I_2}^\dagger,
\end{align}
where $I_2$ and $I_3$ denote the reflected and transmitted idler modes, respectively. Acting on a single idler photon,
\begin{equation}
  |1_I,0_V\rangle \;\longrightarrow\;
  r\,|1,0\rangle_{I_2,I_3}
  + t\,|0,1\rangle_{I_2,I_3}.
\end{equation}

Conditioned on emission from crystal~A, the object transforms the idler into
\begin{equation}
  |\phi_1\rangle
  = r\,|1,0\rangle_{I_2,I_3} + t\,|0,1\rangle_{I_2,I_3},
\end{equation}
while emission from crystal~B, which bypasses the object, prepares the idler in
\begin{equation}
  |\phi_2\rangle
  = |0,1\rangle_{I_2,I_3}.
\end{equation}
These two pure idler states encode the which-crystal hypotheses and have overlap
\begin{equation}
  s := \lvert\langle\phi_1|\phi_2\rangle\rvert = |t| = V,
\end{equation}
where $V$ is the signal singles visibility in the ZWM interferometer.

Unambiguous discrimination between $|\phi_1\rangle$ and $|\phi_2\rangle$ with equal priors requires a three-outcome POVM
$\{\Pi_0,\Pi_1,\Pi_2\}$, corresponding to conclusive identification of $|\phi_1\rangle$, conclusive identification of $|\phi_2\rangle$, and an inconclusive result. Physically, such a POVM can be realized by embedding the two-dimensional idler subspace spanned by $\{|1,0\rangle_{I_2,I_3},|0,1\rangle_{I_2,I_3}\}$ into a three-dimensional single-photon space by introducing a third auxiliary mode $C$ initially in the vacuum state $|0_C\rangle$.

In the single-photon sector of the three modes $(I_2,I_3,C)$ we use the basis
\begin{equation}
\begin{aligned}
  \{\,|100\rangle,\,|010\rangle,\,|001\rangle\,\}_{I_2,I_3,C}
  :=\;
  \{&
     |1,0,0\rangle,\\
     &|0,1,0\rangle,\\
     &|0,0,1\rangle
  \}_{I_2,I_3,C}.
\end{aligned}
\end{equation}

The Ivanovic--Dieks--Peres (IDP) measurement is implemented by a three-mode unitary
$U_{\mathrm{IDP}}$ acting on the extended idler state
$|\phi_j\rangle\otimes|0_C\rangle$. The most general form of $U_{\mathrm{IDP}}$ consistent with error-free discrimination is
\begin{align}
  U_{\mathrm{IDP}}\,|\phi_1,0_C\rangle
  &= \sqrt{p_1}\,|100\rangle + \sqrt{q_1}\,|001\rangle,
  \label{eq:UIDP_phi1}\\
  U_{\mathrm{IDP}}\,|\phi_2,0_C\rangle
  &= \sqrt{p_2}\,|010\rangle + \sqrt{q_2}\,|001\rangle,
  \label{eq:UIDP_phi2}
\end{align}
where $p_j$ and $q_j$ ($j=1,2$) are nonnegative real numbers. In this representation

\begin{itemize}
  \item $p_1$ is the probability that input $|\phi_1\rangle$ produces a click in mode~1 (conclusive ``crystal~A''),
  \item $p_2$ is the probability that input $|\phi_2\rangle$ produces a click in mode~2 (conclusive ``crystal~B''),
  \item $q_1$ and $q_2$ are the probabilities that the respective inputs yield a click in mode~3 (inconclusive outcome).
\end{itemize}

The form of Eqs.~\eqref{eq:UIDP_phi1}–\eqref{eq:UIDP_phi2} enforces the zero-error conditions:
for $|\phi_1\rangle$ there is no amplitude in $|010\rangle$ (the ``B'' detector never fires), and for $|\phi_2\rangle$ there is no amplitude in $|100\rangle$ (the ``A'' detector never fires). Unitarity of $U_{\mathrm{IDP}}$ implies that the output states are normalized,
\begin{equation}
  p_1 + q_1 = 1,\qquad
  p_2 + q_2 = 1.
\end{equation}

Because $U_{\mathrm{IDP}}$ is unitary, it preserves inner products:
\begin{equation}
  \langle\phi_1|\phi_2\rangle
  = \langle\phi_1,0_C|\phi_2,0_C\rangle
  = \langle\mathrm{out}_1|\mathrm{out}_2\rangle,
\end{equation}
where $|\mathrm{out}_j\rangle = U_{\mathrm{IDP}}|\phi_j,0_C\rangle$. Using
Eqs.~\eqref{eq:UIDP_phi1}–\eqref{eq:UIDP_phi2} and the orthogonality
$\langle100|010\rangle=\langle100|001\rangle=\langle010|001\rangle=0$,
$\langle001|001\rangle=1$, we find
\begin{equation}
  \langle\mathrm{out}_1|\mathrm{out}_2\rangle
  = \sqrt{q_1 q_2},
\end{equation}
so that
\begin{equation}
  s = |\langle\phi_1|\phi_2\rangle|
  = \sqrt{q_1 q_2}.
\end{equation}

For equal a priori probabilities and a symmetric optimal strategy, one chooses
\begin{equation}
  q_1 = q_2 = q,\qquad p_1 = p_2 = 1-q,
\end{equation}
so that
\begin{equation}
  q = |\langle\phi_1|\phi_2\rangle| = s.
\end{equation}
The average success probability of the IDP measurement is then
\begin{equation}
  P_S^{\mathrm{opt}}
  = \tfrac{1}{2}p_1 + \tfrac{1}{2}p_2
  = 1 - s,
\end{equation}
and the optimal inconclusive probability is
\begin{equation}
  P_I^{\mathrm{opt}} = 1 - P_S^{\mathrm{opt}} = s
  = |\langle\phi_1|\phi_2\rangle|.
\end{equation}

In the ZWM interferometer, $s = |\langle\phi_1|\phi_2\rangle| = |t| = V$, so the IDP theory yields
\begin{equation}
  P_I^{\mathrm{opt}} = V,\qquad
  P_S^{\mathrm{opt}} = 1 - V.
\end{equation}
Thus the conventional ZWM visibility is exactly the optimal failure probability of an error-free IDP measurement on the idler, while the which-path distinguishability is identified with the conclusive success probability. This provides the operational identity
\begin{equation}
  D^2 + \bigl(P_I^{\mathrm{opt}}\bigr)^2 = 1
\end{equation}
for pure marker states, and establishes the interpretation $V = P_I^{\mathrm{opt}}$ used in the main text.

\section{General two-path interferometer with which-path markers}
\label{seven}

In this Appendix we provide a general derivation of the relation
\begin{equation}
D^2 + \big(P_I^{\mathrm{opt}}\big)^2 = 1
\end{equation}
for a balanced two-path interferometer \cite{mohseni} in the single-excitation regime
with pure which-path markers. We allow arbitrary \emph{a priori}
probabilities $\eta_1,\eta_2$ and show how visibility, distinguishability,
and optimal IDP failure probability are related.

\subsection{A. Joint state and reduced density matrix}

Let $S$ denote the interfering system (paths 1 and 2) and $M$ the
marker system (polarization, spatial mode, idler, etc.). Conditioned
on path $j\in\{1,2\}$, the marker is prepared in a pure state
$|\mu_j\rangle$, and the probabilities of the two alternatives are
$\eta_1,\eta_2$ with $\eta_1+\eta_2=1$.

A convenient pure joint state is
\begin{equation}
|\Psi\rangle =
\sqrt{\eta_1}\,|1,0\rangle_S\otimes|\mu_1\rangle_M
+ \sqrt{\eta_2}\,e^{i\phi}\,|0,1\rangle_S\otimes|\mu_2\rangle_M,
\label{eq:psi_general_appendix}
\end{equation}
where $\phi$ is a controllable phase. Define the marker overlap
\begin{equation}
\gamma := \langle\mu_1|\mu_2\rangle,\qquad |\gamma|\le1.
\end{equation}

The reduced state of the system $S$ is
\begin{equation}
\rho_S = \mathrm{Tr}_M\left(|\Psi\rangle\langle\Psi|\right).
\end{equation}
Using Eq.~\eqref{eq:psi_general_appendix}, we compute
\begin{align}
|\Psi\rangle\langle\Psi|
&=
\eta_1\,|1,0\rangle\langle1,0|\otimes|\mu_1\rangle\langle\mu_1|
+ \eta_2\,|0,1\rangle\langle0,1|\otimes|\mu_2\rangle\langle\mu_2|
\nonumber\\
&\quad
+ \sqrt{\eta_1\eta_2}e^{i\phi}|0,1\rangle\langle1,0|\otimes|\mu_2\rangle\langle\mu_1|
\nonumber\\
&\quad
+ \sqrt{\eta_1\eta_2}e^{-i\phi}|1,0\rangle\langle0,1|\otimes|\mu_1\rangle\langle\mu_2|.
\end{align}
Taking the partial trace over the meter system $M$ gives
\begin{equation}
\mathrm{Tr}_M(|\mu_1\rangle\langle\mu_1|) = 
\mathrm{Tr}_M(|\mu_2\rangle\langle\mu_2|) = 1,
\end{equation}
since $|\mu_j\rangle$ are normalized pure states of $M$ and their projectors have unit trace.  The result is a scalar, not an operator.

\begin{equation}
\mathrm{Tr}_M(|\mu_2\rangle\langle\mu_1|)
= \langle\mu_1|\mu_2\rangle = \gamma.
\end{equation}
Thus
\begin{align}
\rho_S
&=
\eta_1\,|1,0\rangle\langle1,0|
+ \eta_2\,|0,1\rangle\langle0,1|
\nonumber\\
&\quad
+ \sqrt{\eta_1\eta_2}e^{i\phi}\gamma
\,|0,1\rangle\langle1,0|
+ \sqrt{\eta_1\eta_2}e^{-i\phi}\gamma^*
\,|1,0\rangle\langle0,1|.
\end{align}
In the basis $\{|1,0\rangle_S,|0,1\rangle_S\}$,
\begin{equation}
\rho_S =
\begin{pmatrix}
\eta_1 & \sqrt{\eta_1\eta_2}e^{-i\phi}\gamma^*\\[4pt]
\sqrt{\eta_1\eta_2}e^{i\phi}\gamma & \eta_2
\end{pmatrix}.
\label{eq:rhoS_general_appendix}
\end{equation}

\subsection{B. Fringe visibility}

For a balanced output beam splitter, the intensity at one output port
varies with the phase as
\begin{equation}
I(\varphi) = I_0\left[1 + V\cos(\varphi+\varphi_0)\right],
\end{equation}
where $V$ is the fringe visibility and $\varphi_0$ is a phase offset.
In the single-photon regime, the visibility is related to the magnitude
of the coherence (off-diagonal element) of $\rho_S$ as
\begin{equation}
V = \frac{2|\rho_{01}|}{\rho_{00}+\rho_{11}}.
\end{equation}
Since $\rho_{00}+\rho_{11} = \eta_1+\eta_2=1$, we have
\begin{equation}
V = 2|\rho_{01}|.
\end{equation}
From Eq.~\eqref{eq:rhoS_general_appendix},
\begin{equation}
\rho_{01} = \sqrt{\eta_1\eta_2}e^{-i\phi}\gamma^*,
\end{equation}
so
\begin{equation}
|\rho_{01}| = \sqrt{\eta_1\eta_2}|\gamma|.
\end{equation}
Hence
\begin{equation}
V = 2\sqrt{\eta_1\eta_2}\,|\gamma|.
\label{eq:V_general_appendix}
\end{equation}

For \emph{equal priors} $\eta_1=\eta_2=\tfrac12$, this reduces to
\begin{equation}
V = |\gamma| = |\langle\mu_1|\mu_2\rangle|.
\end{equation}

\subsection{C. Distinguishability}

For two pure alternatives $|\mu_1\rangle,|\mu_2\rangle$ with priors
$\eta_1,\eta_2$, a suitable measure of which-path distinguishability is
\begin{equation}
D = \sqrt{1 - 4\eta_1\eta_2|\gamma|^2}.
\label{eq:D_general_appendix}
\end{equation}
With Eq.~\eqref{eq:V_general_appendix}, we then have
\begin{equation}
D^2 + V^2
= \big(1 - 4\eta_1\eta_2|\gamma|^2\big)
+ 4\eta_1\eta_2|\gamma|^2
= 1.
\end{equation}
Thus, in this pure-state setting,
\begin{equation}
D^2 + V^2 = 1
\end{equation}
holds for arbitrary priors. For equal priors $\eta_1=\eta_2=\tfrac12$,
Eq.~\eqref{eq:D_general_appendix} simplifies to
\begin{equation}
D = \sqrt{1-|\gamma|^2}
= \sqrt{1-V^2}.
\end{equation}

\subsection{D. IDP optimal failure probability}

Unambiguous discrimination between the two pure marker states
$|\mu_1\rangle,|\mu_2\rangle$ with priors $\eta_1,\eta_2$ is described by a
three-outcome POVM $\{\Pi_1,\Pi_2,\Pi_{\rm inc}\}$ satisfying the
zero-error constraints
$\Tr(\Pi_1|\mu_2\rangle\langle\mu_2|)=0$ and
$\Tr(\Pi_2|\mu_1\rangle\langle\mu_1|)=0$.
The optimal inconclusive probability is, in general, piecewise.
In the symmetric-IDP regime one has~\cite{Ivanovic,Dieks,Peres,Barnett}
\begin{equation}
P_{\rm inc}^{\mathrm{opt}} = 2\sqrt{\eta_1\eta_2}\,|\gamma|,
\label{eq:PIopt_symmetric}
\end{equation}
and therefore, comparing with Eq.~\eqref{eq:V_general_appendix},
\begin{equation}
P_{\rm inc}^{\mathrm{opt}} = V
\qquad
\text{(symmetric-IDP branch)}.
\end{equation}

\paragraph*{Remark on unequal priors.}
For unequal a priori probabilities, the optimal unambiguous state
discrimination (USD) strategy is in general \emph{piecewise}.
The expression \eqref{eq:PIopt_symmetric} holds in the symmetric-IDP
regime, which includes the equal-prior case
$\eta_1=\eta_2=\tfrac{1}{2}$ relevant to equal pumping in the ZWM
interferometer. Outside this regime, the optimal solution changes
branch, and one of the conclusive outcomes may be suppressed.
Since the present work focuses on the equal-pump induced-coherence
scenario, we restrict attention to the symmetric-IDP branch.

\subsection{E. Remarks on unequal priors and mixed markers}

The above derivation assumes that unambiguous discrimination is
possible (i.e.\ the supports of the two states are not nested) and that
we are in the regime where the symmetric IDP solution applies. In
extreme cases of highly asymmetric priors, the optimal strategy may
change branch, but for the equal-pump ZWM scenario of interest
($\eta_1=\eta_2=\tfrac12$) the expression
$P_I^{\mathrm{opt}} = |\langle\mu_1|\mu_2\rangle|$ holds directly.

For mixed marker states $\rho_1,\rho_2$, the visibility is bounded by
the fidelity $F(\rho_1,\rho_2)$,
\begin{equation}
V \le F(\rho_1,\rho_2),
\end{equation}
and the structure of optimal unambiguous discrimination becomes more
involved, depending on the supports of $\rho_1$ and $\rho_2$. A full
mixed-state analysis is beyond the scope of this work. In the
low-gain, pure-marker regime relevant to the ZWM interferometer,
Eqs.~\eqref{eq:V_general_appendix}--\eqref{eq:D_general_appendix} and
the operational identity $D^2+(P_I^{\mathrm{opt}})^2=1$ remain exact.

\section{Experimental protocol}\label{eight}
Previous unambiguous path discrimination in Mach–Zehnder interferometers \cite{Len2018IDP,Clarke2001IDP}, where USD identifies orthogonal signal paths via an auxiliary degree of freedom. The present work applies unambiguous state discrimination to a fundamentally different task. In the low-gain induced-coherence interferometer, the idler modes $I$ are deliberately aligned to be path-identical, rendering which-path information ill-defined for the idler photon. The only meaningful alternative encoded in the idler is therefore the source of the photon pair, namely which nonlinear crystal emitted it. Emission from crystal A and crystal B prepares two distinct but generally nonorthogonal conditional idler states. Unambiguous state discrimination on the idler thus constitutes a zero-error retrodictive measurement of the emitting crystal, rather than a which-path measurement. This distinction is essential: USD here addresses source identification in an induced-coherence geometry, not path discrimination in an interferometer.

In the IDP framework, the three-outcome measurement is described by the POVM \cite{mohseni,Barnett}
$\{\Pi_0,\Pi_1,\Pi_2\}$, where $\Pi_1$ and $\Pi_2$ correspond to conclusive
identification of $|\phi_1\rangle$ and $|\phi_2\rangle$, and $\Pi_0$
represents the inconclusive outcome. For a state $\rho$ entering the
measurement device, the outcome probabilities follow from
\begin{equation}
P_i = \mathrm{Tr}(\rho\,\Pi_i), \qquad i = 0, 1, 2.
\end{equation}
For equal priors $\rho=\frac{1}{2}(|\phi_1\rangle\langle\phi_1|
+|\phi_2\rangle\langle\phi_2|)$, the Ivanovic--Dieks--Peres theorem gives
\begin{equation}
P_I^{\mathrm{opt}} = |\langle\phi_1|\phi_2\rangle|.
\end{equation}
In the ZWM interferometer, this overlap also sets the signal visibility,
$V = |\langle\phi_1|\phi_2\rangle|$, and therefore
\begin{equation}
P_I^{\mathrm{opt}} = V.
\end{equation}
Experimentally, the probabilities are estimated from the detected photon
counts $N_1,N_2,N_0$ through relative frequencies
\begin{equation}
P_S^{\mathrm{opt}}=\frac{N_1+N_2}{N_1+N_2+N_0}, \quad
P_I^{\mathrm{opt}}=\frac{N_0}{N_1+N_2+N_0}.
\end{equation}

\section{Mixed idler marker states with thermal noise}
\label{nine}

In the main text we worked in the ideal low-gain regime with pure idler marker states, for which the overlap
$|\langle\phi_1|\phi_2\rangle|$ fixes both the single-photon
visibility and the optimal IDP failure probability,
$V = |\langle\phi_1|\phi_2\rangle| = P_I^{\mathrm{opt}}$.
In this Appendix we outline how the picture is modified when thermal background photons are injected into the idler path $S$. The which-crystal markers become mixed states. We work in the Schr\"odinger picture at the level of density operators, and make use of Uhlmann fidelity rather than repeating the full Heisenberg calculation \cite{boyd,giri1}.

\subsection{A. Thermal environment and conditional idler states}
We model the object $S$ in the common idler path between crystals~A and~B as a lossless beam splitter that couples the idler mode to a thermal background
mode.  Denote the corresponding annihilation operators by
\begin{equation}
  \hat a_I \;\, \text{(idler mode)}, \qquad
  \hat a_4 \;\, \text{(environment mode)}.
\end{equation}
The environment is taken to be in a single-mode thermal state with mean photon
number $N_B = \langle \hat a_4^\dagger \hat a_4 \rangle$, described by
\begin{equation}
  \rho_{\mathrm{th}}(N_B)
  = \sum_{n=0}^\infty p_n\,|n\rangle\langle n|,
  \qquad
  p_n = \frac{N_B^n}{(1+N_B)^{n+1}}.
\end{equation}

The object $S$ is represented by a beam-splitter unitary $U_S$ with
amplitude transmittance $t_o$ and reflectivity $r_o$,
$|t_o|^2+|r_o|^2=1$, acting as
\begin{equation}
  U_S^\dagger
  \begin{pmatrix}
    \hat a_{I_2}\\[2pt] \hat a_{I_3}
  \end{pmatrix}
  U_S
  =
  \begin{pmatrix}
    r_o & t_o\\[2pt]
    t_o & -r_o
  \end{pmatrix}
  \begin{pmatrix}
    \hat a_I\\[2pt] \hat a_4
  \end{pmatrix},
\end{equation}
where $I_3$ is the transmitted idler mode that seeds crystal~B (and is ultimately delivered to the IDP apparatus), and $I_2$ is the loss port. In the single-pair regime, the relevant sector is that of at most one idler photon plus an arbitrary thermal population in mode~4.

We now define the conditional idler marker states \emph{after} the
object $S$, given that either crystal~A or crystal~B emitted the photon pair. We treat the thermal environment in the same way for both hypotheses, so that both markers are defined on the same effective output mode $I_3$.

\paragraph*{Emission from crystal A.}
Conditioned on emission in crystal~A, the idler is in a one-photon state
in mode $I$ before the object, while the thermal environment is in
$\rho_{\mathrm{th}}(N_B)$. The joint input state is therefore
\begin{equation}
  \rho^{(A)}_{\mathrm{in}}
  = |1_I\rangle\langle 1_I|
    \otimes \rho_{\mathrm{th}}(N_B).
\end{equation}
After the beam splitter $U_S$, the global state of modes $I_2$, $I_3$
and $4$ is
\begin{equation}
  \rho^{(A)}_{\mathrm{out}}
  = U_S\,\rho^{(A)}_{\mathrm{in}}\,U_S^\dagger.
\end{equation}
The idler marker associated with ``crystal~A fired'' is defined as the
reduced state of the transmitted idler mode $I_3$, obtained by tracing
over the loss port $I_2$ and the remaining environment mode~4:
\begin{equation}
  \rho_A
  =
  \mathrm{Tr}_{I_2,4}\big(\rho^{(A)}_{\mathrm{out}}\big).
  \label{eq:rhoA_mixed_def}
\end{equation}
Because $U_S$ creates correlations between $I_3$ and the traced modes,
and because the thermal state is a classical mixture, $\rho_A$ is in
general a \emph{mixed} state on the idler Hilbert space. Physically,
this mixedness represents the fact that some which-path information has leaked into unobserved degrees of freedom (loss port + thermal bath).

\paragraph*{Emission from crystal B.}
Conditioned on emission in crystal~B, the signal photon travels along
path~2, while the idler photon is injected into the common idler path
that ultimately defines mode~$I_3$ at the IDP apparatus. To treat both
hypotheses on the same footing, we also let the idler interact with the
same thermal environment via $U_S$, but now with a possibly different
effective input mode or coupling. For concreteness, we take the joint
input state under hypothesis~B to be
\begin{equation}
  \rho^{(B)}_{\mathrm{in}}
  = |\psi_B\rangle\langle\psi_B|
    \otimes \rho_{\mathrm{th}}(N_B),
\end{equation}
where $|\psi_B\rangle$ is a (generally different) single-mode pure state
of the idler that reaches the object when crystal~B emits. After the
$U_S$ we have
\begin{equation}
  \rho^{(B)}_{\mathrm{out}}
  = U_S\,\rho^{(B)}_{\mathrm{in}}\,U_S^\dagger,
\end{equation}
and the corresponding idler marker is
\begin{equation}
  \rho_B
  =
  \mathrm{Tr}_{I_2,4}\big(\rho^{(B)}_{\mathrm{out}}\big).
  \label{eq:rhoB_mixed_def}
\end{equation}
Again, the partial trace over the loss port and thermal mode produces a
mixed state $\rho_B$ on mode~$I_3$ in general. In this way, both
which-crystal alternatives are represented by density operators
$\rho_A$ and $\rho_B$ on the same idler output mode, and both include
the effect of thermal background photons.

We emphasize that the thermal photons are not entangled with the idler
in any nonclassical sense; rather, the beam splitter generates
mode entanglement between $(I_2,I_3)$ and the environment, and tracing
over the unobserved modes converts the conditional idler state into a
mixed density operator.

\subsection{B. Purifications, visibility, and Uhlmann fidelity}

To relate the single-photon fringe visibility in the signal arm to the
conditional idler markers $\rho_A$ and $\rho_B$, it is convenient to work with
explicit purifications.  Let $M$ denote the transmitted idler mode $I_3$, and
let $E$ denote an abstract environment that collects the loss port $I_2$, the
thermal mode~4, and any additional vacuum modes.  By definition, each mixed
idler state admits a purification on $ME$, so we introduce pure states
$|\Psi_A\rangle_{ME}$ and $|\Psi_B\rangle_{ME}$ satisfying
\begin{equation}
  \rho_A = \mathrm{Tr}_E\big(|\Psi_A\rangle\langle\Psi_A|\big), \qquad
  \rho_B = \mathrm{Tr}_E\big(|\Psi_B\rangle\langle\Psi_B|\big).
\end{equation}
In the low-gain regime with equal pumping of the two crystals, the joint pure
state of the signal-path qubit $S$ and the idler--environment system $ME$ is
\begin{equation}
  |\Phi\rangle_{SME}
  =
  \frac{1}{\sqrt{2}}\Big(
    |1\rangle_S\otimes|\Psi_A\rangle_{ME}
    + e^{i\phi}\,|2\rangle_S\otimes|\Psi_B\rangle_{ME}
  \Big),
  \label{eq:Phi_thermal_mixed}
\end{equation}
where $|1\rangle_S$ and $|2\rangle_S$ label the two signal paths and $\phi$ is
a controllable phase.

The reduced state of the interfering signal system is obtained by
tracing out the idler and environment,
\begin{equation}
  \rho_S = \mathrm{Tr}_{ME}\big(|\Phi\rangle\langle\Phi|\big).
\end{equation}
A straightforward calculation in the basis $\{|1\rangle_S,|2\rangle_S\}$
gives
\begin{equation}
  \rho_S
  =
  \begin{pmatrix}
    \tfrac12
    & \tfrac12\,e^{-i\phi}\langle \Psi_B | \Psi_A \rangle \\[4pt]
    \tfrac12\,e^{i\phi}\langle \Psi_A | \Psi_B \rangle
    & \tfrac12
  \end{pmatrix}.
\end{equation}

The off-diagonal element $(\rho_S)_{12}$ determines the visibility of
the single-photon interference fringes at the output of the balanced
signal beam splitter. For a symmetric two-path interferometer one finds
\begin{equation}
  V
  = \frac{2|(\rho_S)_{12}|}{\rho_{11}+\rho_{22}}
  = 2\left|\frac12\langle\Psi_B|\Psi_A\rangle\right|
  = |\langle\Psi_A|\Psi_B\rangle|.
  \label{eq:V_overlap_purif_mixed}
\end{equation}

The Uhlmann fidelity between the mixed idler markers is defined as
\begin{equation}
  F(\rho_A,\rho_B)
  :=
  \max_{|\Psi_A\rangle,|\Psi_B\rangle}
  |\langle\Psi_A|\Psi_B\rangle|,
  \label{eq:fidelity_def}
\end{equation}
where the maximization is over all purifications of $\rho_A$ and
$\rho_B$. Comparing Eqs.~\eqref{eq:V_overlap_purif_mixed}
and~\eqref{eq:fidelity_def}, we immediately obtain the general bound
\begin{equation}
  V \;\le\; F(\rho_A,\rho_B),
  \label{eq:V_leq_F_mixed}
\end{equation}
with equality attainable for a suitable choice of coupling between the
idler and its environment. In the pure-marker limit
$\rho_A = |\phi_1\rangle\langle\phi_1|$ and
$\rho_B = |\phi_2\rangle\langle\phi_2|$, we can choose
$|\Psi_A\rangle = |\phi_1\rangle\otimes|0_E\rangle$ and 
$|\Psi_B\rangle = |\phi_2\rangle\otimes|0_E\rangle$, so that
$F(\rho_A,\rho_B)=|\langle\phi_1|\phi_2\rangle|$ and
Eq.~\eqref{eq:V_leq_F_mixed} reproduces the ideal relation
$V = |\langle\phi_1|\phi_2\rangle|$ used in the main text.

\subsection{C. Optimal IDP failure probability for mixed markers}

We now connect the mixed idler markers $\rho_A$ and $\rho_B$ to optimal
unambiguous state discrimination. Consider any three-outcome POVM
$\{\Pi_A,\Pi_B,\Pi_0\}$ on the idler mode $I_3$, where $\Pi_A$ and
$\Pi_B$ correspond to conclusive identification of ``crystal~A'' and
``crystal~B'', respectively, and $\Pi_0$ represents an inconclusive
outcome. For equal a priori probabilities $\eta_A=\eta_B=\tfrac12$, the
average failure probability is
\begin{equation}
  P_I = \frac12\,\Tr(\rho_A\Pi_0) + \frac12\,\Tr(\rho_B\Pi_0),
\end{equation}
subject to the zero-error constraints
\begin{equation}
  \Tr(\rho_A\Pi_B) = 0,\qquad
  \Tr(\rho_B\Pi_A) = 0.
\end{equation}
The optimal IDP failure probability $P_I^{\mathrm{opt}}$ is obtained by
minimizing $P_I$ over all such POVMs. For two mixed states, a general
lower bound is known in terms of the Uhlmann fidelity
$F(\rho_A,\rho_B)$~\cite{Barnett,Herzog2005MixedUSD,Uhlmann76}:
\begin{equation}
  P_I^{\mathrm{opt}}
  \;\ge\; F(\rho_A,\rho_B),
  \label{eq:PI_geq_F_mixed}
\end{equation}
with equality attained for pure states and for certain special classes
of mixed states. In particular, for
$\rho_A = |\phi_1\rangle\langle\phi_1|$ and
$\rho_B = |\phi_2\rangle\langle\phi_2|$ one recovers the standard IDP
result $P_I^{\mathrm{opt}} = |\langle\phi_1|\phi_2\rangle|$.

Combining Eqs.~\eqref{eq:V_leq_F_mixed} and~\eqref{eq:PI_geq_F_mixed}
for the thermal-seeded ZWM markers defined in
Eqs.~\eqref{eq:rhoA_mixed_def} and~\eqref{eq:rhoB_mixed_def}, we obtain
the mixed-state operational duality inequality
\begin{equation}
  V \;\le\; F(\rho_A,\rho_B)
    \;\le\; P_I^{\mathrm{opt}}.
  \label{eq:V_leq_PI_mixed_final}
\end{equation}
In words: in the presence of thermal background photons, the observed
signal singles visibility is always bounded above by the Uhlmann
fidelity between the mixed idler markers, which in turn is bounded above
by the optimal inconclusive probability of any unambiguous
which-crystal measurement on the idler. In the ideal pure-marker limit
(realized in the absence of thermal noise and loss), the inequalities in
Eq.~\eqref{eq:V_leq_PI_mixed_final} collapse to the equality
\begin{equation}
  V = F(\rho_A,\rho_B) = P_I^{\mathrm{opt}},
\end{equation}
recovering the main result of the text,
$V = P_I^{\mathrm{opt}}$, and showing that the standard
complementarity relation $D^2+V^2=1$ can be written as
$D^2+(P_I^{\mathrm{opt}})^2=1$ for the ZWM interferometer in the
low-noise regime.

\end{document}